\documentclass[prb,superscriptaddress,twocolumn]{revtex4-1}
\pdfoutput=1
\usepackage[colorlinks=true,urlcolor=blue]{hyperref}
\usepackage[normalem]{ulem}
\usepackage[utf8]{inputenc}
\usepackage[english]{babel}
\usepackage[T1]{fontenc}
\usepackage{mathtools}
\usepackage{graphicx}
\usepackage{placeins}
\usepackage{amsfonts}
\usepackage{stmaryrd}
\usepackage{makecell}
\usepackage{rotating}
\usepackage{multirow}
\usepackage{booktabs}
\usepackage{arydshln}
\usepackage{amsmath}
\usepackage{xcolor}
\usepackage{amsthm}
\usepackage{xcolor}
\usepackage{bm}
\usepackage[normalem]{ulem}

\definecolor{red}{rgb}{0.75,0,0}
\definecolor{blue}{rgb}{0,0,0.75}
\definecolor{green}{rgb}{0,0.5,0}
\definecolor{airforceblue}{rgb}{0.36, 0.54, 0.66}

\newcommand{\traceless}[1]{\left\llbracket #1 \right\rrbracket}

\DeclareMathOperator{\Arg}{Arg}

\begin{document}

\title{Epithelia are multiscale active liquid crystals}
\author{Josep-Maria Armengol-Collado}
\thanks{These authors contributed equally}
\affiliation{Instituut-Lorentz, Leiden Institute of Physics, Universiteit Leiden, P.O. Box 9506, 2300 RA Leiden, The Netherlands}
\author{Livio Nicola Carenza}
\thanks{These authors contributed equally}
\affiliation{Instituut-Lorentz, Leiden Institute of Physics, Universiteit Leiden, P.O. Box 9506, 2300 RA Leiden, The Netherlands}
\author{Julia Eckert}
\thanks{These authors contributed equally}
\affiliation{Instituut-Lorentz, Leiden Institute of Physics, Universiteit Leiden, P.O. Box 9506, 2300 RA Leiden, The Netherlands}
\affiliation{Physics of Life Processes, Leiden Institute of Physics, Universiteit Leiden, P.O. Box 9506, 2300 RA Leiden, The Netherlands}
\author{Dimitris Krommydas}
\thanks{These authors contributed equally}
\affiliation{Instituut-Lorentz, Leiden Institute of Physics, Universiteit Leiden, P.O. Box 9506, 2300 RA Leiden, The Netherlands}
\author{Luca Giomi}
\email{giomi@lorentz.leidenuniv.nl}
\affiliation{Instituut-Lorentz, Leiden Institute of Physics, Universiteit Leiden, P.O. Box 9506, 2300 RA Leiden, The Netherlands}
\date{\today}

\begin{abstract} 
Biological processes such as embryogenesis, wound healing and cancer progression, crucially rely on the ability of epithelial cells to coordinate their mechanical activity over length scales order of magnitudes larger than the typical cellular size~\cite{Friedl2009}. While regulated by various signalling pathways~\cite{DePascalis2017}, it has recently become evident that this behavior can additionally hinge on a minimal toolkit of physical mechanisms, of which liquid crystal order is the most prominent example~\cite{Duclos2016,Saw2017}. Yet, experimental and theoretical studies have given so far inconsistent results in this respect: whereas {\em nematic} order is often invoked in the interpretation of experimental data~\cite{Duclos2016,Saw2017,Kawaguchi2017}, computational models have instead suggested that {\em hexatic} order could serve as a linchpin for collective migration in confluent cell layers~\cite{Li2018,Durand2019,Pasupalak2020}. In this article we resolve this dilemma. Using a combination of {\em in vitro} experiments on Madin-Darby canine kidney cells (MDCK), numerical simulations and analytical work, we demonstrate that both nematic and hexatic order are, in fact, present in epithelial layers, with the former being dominant at large length scales and the latter at small length scales. In MDCK GII cells on uncoated glass, these different types of liquid crystal order crossover at a length scale of the order of ten cell sizes. Our work sheds light on the emergent organization of living matter, provides a new framework for deciphering the structure of epithelia, and paves the way toward a comprehensive and predictive mesoscopic theory of tissues~\cite{Armengol2021}.
\end{abstract}

\maketitle

\begin{figure}[htbp]
\centering 
\includegraphics[width=1.00\columnwidth]{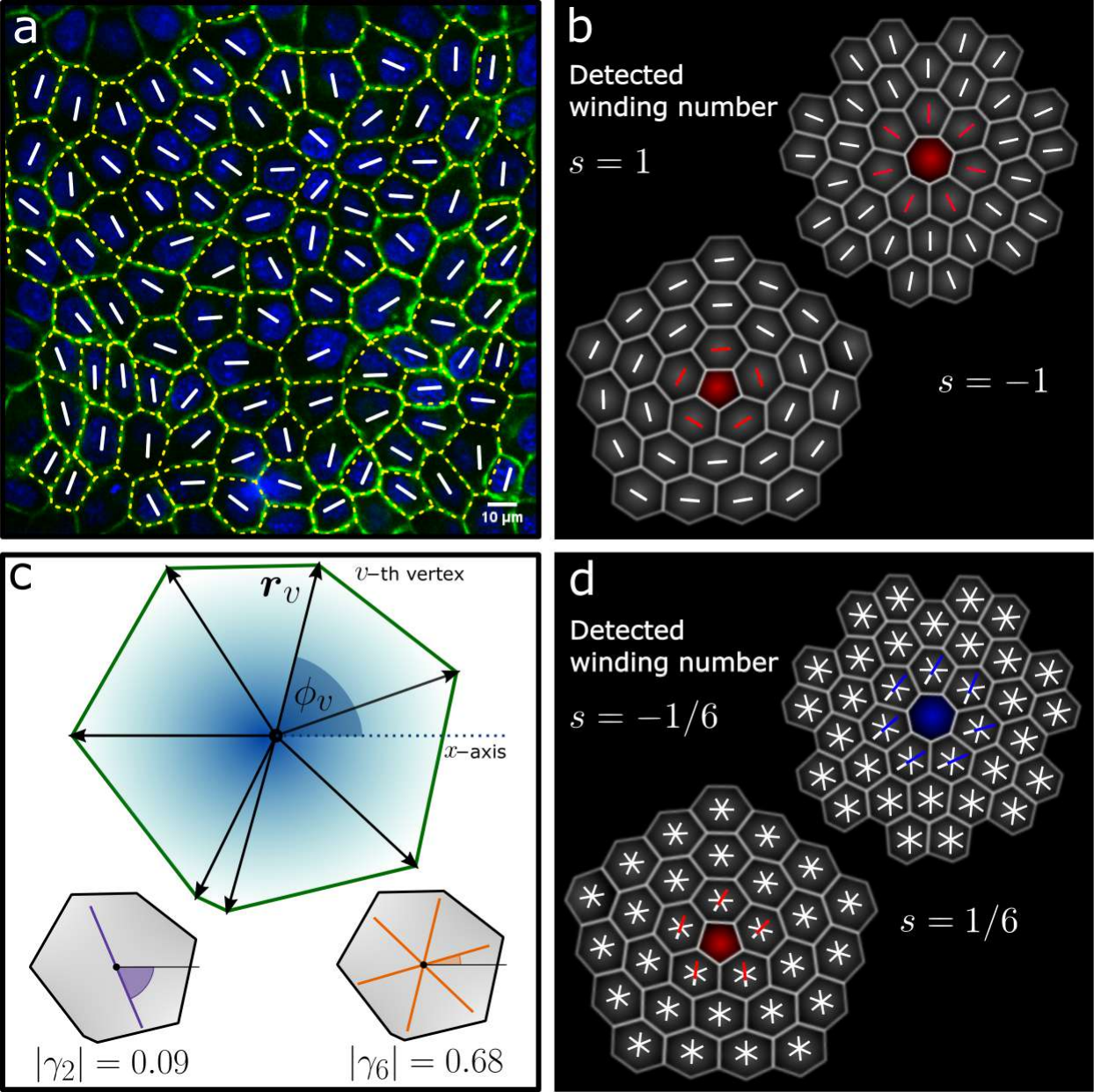}
\caption{\textbf{Orientational order parameter and defect misidentification.} 
\textbf{(a)} In the background: a confocal image of a confluent MDCK GII monolayer (green, E-cadherin and blue, nuclei). The dashed yellow lines in the foreground trace the contour of the cells as identified after image segmentation. The white rods, corresponding to the cells' orientation, have been obtained by computing the shape function $\gamma_{2}$ in Eq.~\eqref{eqn:DefinitionOrderParameter}. \textbf{(b)} Disclinations in hexatics consist of pentagonal (i.e. $s=1/6$) and heptagonal (i.e. $s=-1/6$) cells embedded in an otherwise $6-$fold background. Attempting to detect these elementary defects by tracking the longitudinal direction of the cells (with rods), correctly yields a defect at the center of the clusters, however, because of the mismatch between the $6-$fold symmetry of the configuration and the $2-$fold symmetry of the order parameter, both defects are detected with the incorrect winding number $s=1$. \textbf{(c)} Graphical representation of the $p-$fold shape function in Eq.~\eqref{eqn:DefinitionOrderParameter}, for a generic polygon. The quantities $\bm{r}_{v}$ and $\phi_{v}$ correspond, respectively, to the position of the $v-$th vertex with respect to center of mass of the polygon and its orientation with respect to the $x-$axis. The insets show the phases (black line and red star) and magnitudes of the associated shape functions $\gamma_{2}$ and $\gamma_{6}$. \textbf{(d)} Same defect configurations as in panel \textbf{(b)}, but analyzed via the shape function $\gamma_{6}$. In this case, both the location \emph{and} strength $s=\pm 1/6$ of the defect are correctly identified. One of the legs of the hexatic stars has been colored to facilitate the visualization of the fractional winding number along the closed contour defined by the nearest neighbors of the disclinations.}
\label{Fig:fig1}
\end{figure}

An increasingly large body of evidence suggests that liquid crystal order could lie at the heart of a myriad of cellular processes that are instrumental for life~\cite{Duclos2016,Saw2017,Kawaguchi2017,Balasubramaniam2021}. These include the extrusion of apoptotic cells~\cite{Saw2017}, the development of sharp morphological features, such as tentacles and protrusions in developing embryos~\cite{Maroudas2021,Hoffmann2022}, or the onset of organism-wide cellular flows during morphogenesis~\cite{Streichan2018}. In confluent epithelial layers liquid crystalline order is detected by tracking the longitudinal direction of individual cells. This is most commonly achieved by diagonalizing a rank$-2$ tensor $-$ the so-called structure tensor~\cite{Bigun1987} or equivalently the shape tensor~\cite{Aubouy2003,Asipauskas2003} in case of segmented images $-$ that embodies the geometry of the polygonal cells (Fig.~\ref{Fig:fig1}a). The resulting two-dimensional orientation field is then used to identify topological defects~\cite{Duclos2016,Saw2017,Kawaguchi2017,Balasubramaniam2021}, which in turn provide a fingerprint of the underlying orientational order. In two-dimensional liquid crystals, defects (also known as disclinations), are isolated point singularities in the orientational field and can be classified according to their winding number or ``strength'' $s$, defined as the number of revolutions of the orientation field along an arbitrary contour encircling the defect core~\cite{Chaikin1995}. Because in a liquid crystal with $p-$fold rotational symmetry (i.e. symmetry under rotations by $2\pi/p$) this number must be an integer multiple of $1/p$, defects such as vortices, asters and spirals, for which $s=1$, are a signature of a polar phase (i.e. $p=1$); comet and star-shaped disclinations, whose winding numbers are $s=1/2$ and $s=-1/2$, respectively, are representative of a nematic phase (i.e. $p=2$); whereas $5-$fold and $7-$fold disclinations, with $s=1/6$ and $s=-1/6$ are respectively the elementary topological defects in hexatics (i.e. $p=6$).

Although inferring order from defects represents a consolidated strategy in liquid crystals science since the times of Georges Friedel~\cite{Friedel1922} $-$ who used it to decipher and classify phases such as nematic, cholesteric, and smectic $-$ this specific protocol, based on tracking the cells' longitudinal direction, becomes progressively less reliable as $p$ increases. When applied to a perfect honeycomb lattice, for instance, the protocol leads to the misdetection of a pair of $\pm 1/2$ nematic disclinations (see Extended Data Fig.~E1). This ambiguity originates from the fact that, while regular hexagons are invariant under rotations by $2\pi/6$, the orientation field constructed from the longitudinal direction of hexagonal cells cannot discriminate between the three equivalent directions defined by pairs of opposite vertices. Similarly, in Fig.~\ref{Fig:fig1}b, we show how detecting an elementary hexatic disclination (with strength $s=\pm1/6$) correctly yields a topological defect, but with incorrect winding number $s=1$. 

\begin{figure}[htbp]
\centering 
\includegraphics[width=1.0\columnwidth]{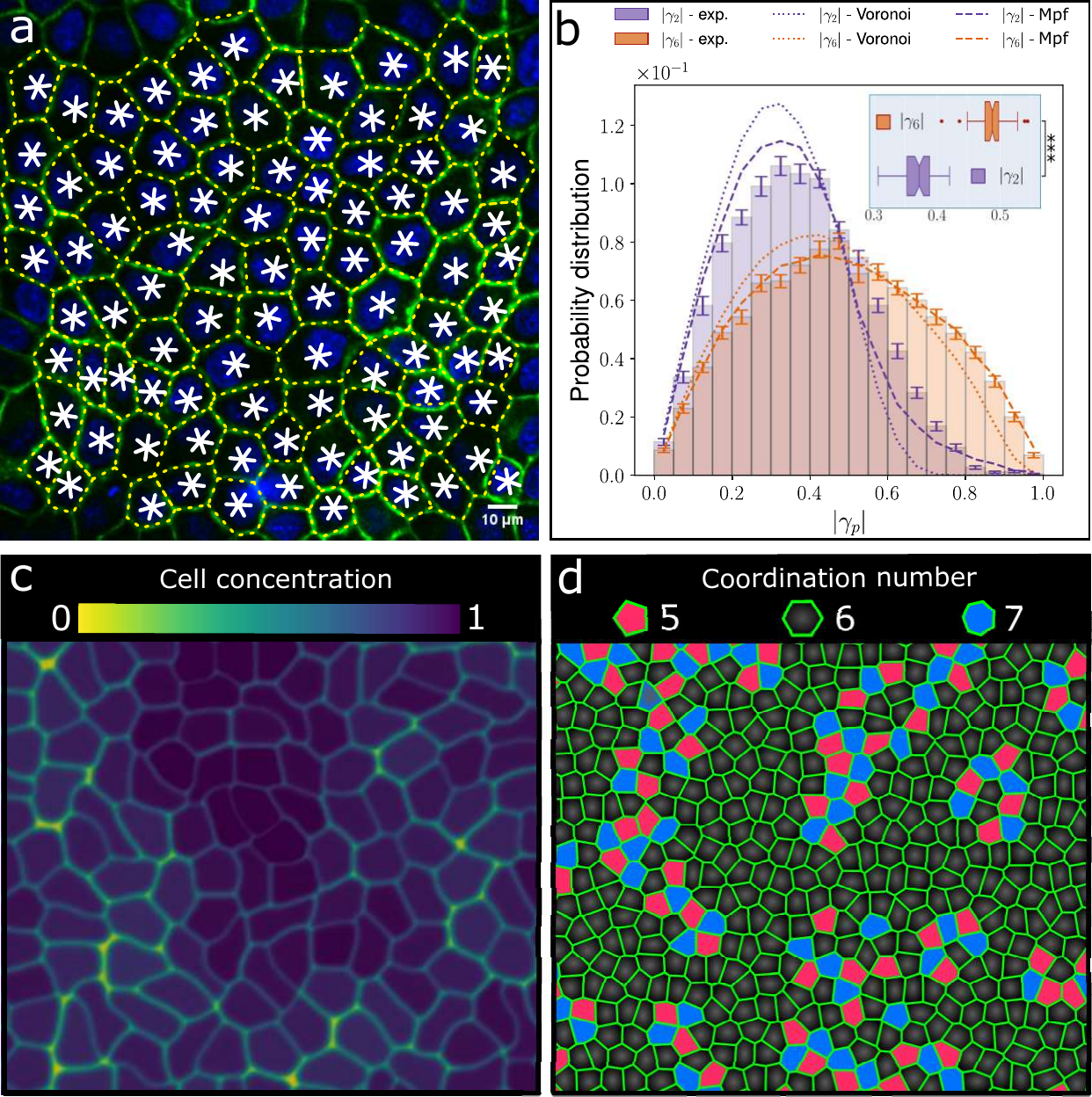}
\caption{\textbf{Symmetry of MDCK cells in confluent monolayers.} 
\textbf{(a)} In the background: same configuration as in Fig.~\ref{Fig:fig1}a. The white stars in the foreground define the $6-$fold orientation of the cells, obtained from the shape function $\gamma_{6}$ as defined in Eq.~\eqref{eqn:DefinitionOrderParameter}. \textbf{(b)}~Probability distribution of the magnitude of $|\gamma_p|$ for $p=2$ (purple) and $p=6$ (orange). Experimental data points are obtained by averaging over $68$ different images, each containing $140 \pm 31$ (mean $\pm$ s.d.) cells. The mean values of the distributions are $\langle |\gamma_2| \rangle = 0.370 \pm 0.030 $ (mean $\pm$ s.d.) and $\langle |\gamma_6| \rangle = 0.49 \pm 0.05 $ (mean $\pm$ s.d.). The boxplot in the inset shows the average of $|\gamma_{2}|$ and $|\gamma_{6}|$ over $9496$ cells across $68$ different samples. These are significantly different with a ${\rm p}-$value of ${\rm p}<10^{-4}$, calculated by using the two-sided Wilcoxon rank sum test. Dashed and dotted lines in the main panel refer to numerical simulations of the multiphase field and Voronoi models. The model parameters have been chosen to minimize the $L_2-$ distance of $|\gamma_6|$ model data from their experimental counterpart. \textbf{(c)}~Contour plot of the local cell concentration of a multiphase-field simulation with $360$ cells in a magnified region of the simulation box, showing approximately one third of the system. Darker regions correspond to areas dense with cells and lighter regions to areas where cells are sparser. Identification of vertices of the model cells is carried out through image segmentation. \textbf{(d)}~Zoom into a typical configuration obtained from numerical simulations of the Voronoi model. Cells with $5$ ($7$) neighbors are highlighted in red (blue). All other topologically neutral cells are plotted in black (see Methods for details).}
\label{Fig:fig2}
\end{figure}

To overcome this limitation, we introduce a generalized rank$-p$ shape tensor, able to capture arbitrary $p-$fold rotational symmetries, with $p$ any natural number. Given the polygonal contour of a cell, whose $V$ vertices have coordinates $\bm{r}_{v}$ with respect to the cell's center (Fig.~\ref{Fig:fig1}c), our generalized shape tensor can be defined as
\begin{equation}
\bm{G}_{p} = \frac{1}{\Delta_p} \bigg\llbracket\sum_{v=1}^{V} \underbrace{\bm{r}_{v} \otimes \bm{r}_{v} \otimes \dots \otimes \bm{r}_{v}}_{p\,{\rm times}}\bigg\rrbracket\;,
\label{eqn:pRankTensor}
\end{equation}
where $\Delta_{p} = \sum_{v=1}^{V} |\bm{r}_v|^{p}$, and the operator $\traceless{\cdots}$ has the effect of rendering its argument symmetric and traceless~\footnote{For tensors whose rank is higher than two, the property of being traceless implies that contracting any two indices of the tensor yields zero.}. For $p=2$, Eq.~\eqref{eqn:pRankTensor} gives, up to a normalization constant, the traceless part of the standard rank$-2$ shape tensor~\cite{Aubouy2003,Asipauskas2003}. Regardless of its rank, the tensor $\bm{G}_{p}$ has only two linearly independent components in two dimensions~\cite{Giomi2021a,Giomi2021b}, from which one can extract information about the cells' shape and orientation. In particular, using a generalization of the spectral theorem to tensors with arbitrary rank~\cite{Virga2015}, one can show (Supplementary Information, Sec.~S2) that all elements of $\bm{G}_{p}$ are proportional to either the real or the imaginary part of the complex function
\begin{equation}
\gamma_{p} = \frac{1}{\Delta_{p}} \sum_{v=1}^{V} |\bm{r}_{v}|^{p} e^{ip\phi_{v}}\;,
\label{eqn:DefinitionOrderParameter}
\end{equation}
where $\phi_{v}$ is the angle between the $v-$th vertex of a given cell and the $x-$axis (Fig.~\ref{Fig:fig1}c). In the following, we will refer to $\gamma_{p}$ as {\em shape function}. Its phase -- i.e. $\vartheta=\Arg(\gamma_{p})/p$ -- corresponds to the $p-$fold orientation of the whole cell with respect to the horizontal direction. In practice, this is equivalent to the inclination of a $p-$legged star located at the cell's center and oriented in a way that maximizes the probability of finding a vertex in the direction of either one of the legs. The shape function $\gamma_{p}$ can be viewed as a generalization of the traditional $p-$fold orientational order parameter $\psi_{p}=\exp (ip\vartheta)$~\cite{Halperin1978,Nelson1979} to irregular and unequal shapes and, unlike the latter, has the advantage of quantifying {\em both} the shapes' orientation -- via its phase $0 \le \vartheta < 2\pi/p$ -- and regularity -- via its magnitude $0\le |\gamma_{p}| \le 1$ (see insets of Fig.~\ref{Fig:fig1}c and Extended Data Fig.~E2). Moreover, when used to detect hexatic disclinations, $\gamma_{6}$ yields the correct winding number $s=\pm 1/6$ (Fig.~\ref{Fig:fig1}b,d). Further details about the shape function $\gamma_{p}$, including a benchmark analysis on systems of melting Brownian particles~\cite{Bernard2011,Digregorio2018}, can be found in the Supplementary Information.

\begin{figure*}[hbtp]
\centering 
\includegraphics[width=1.0\textwidth]{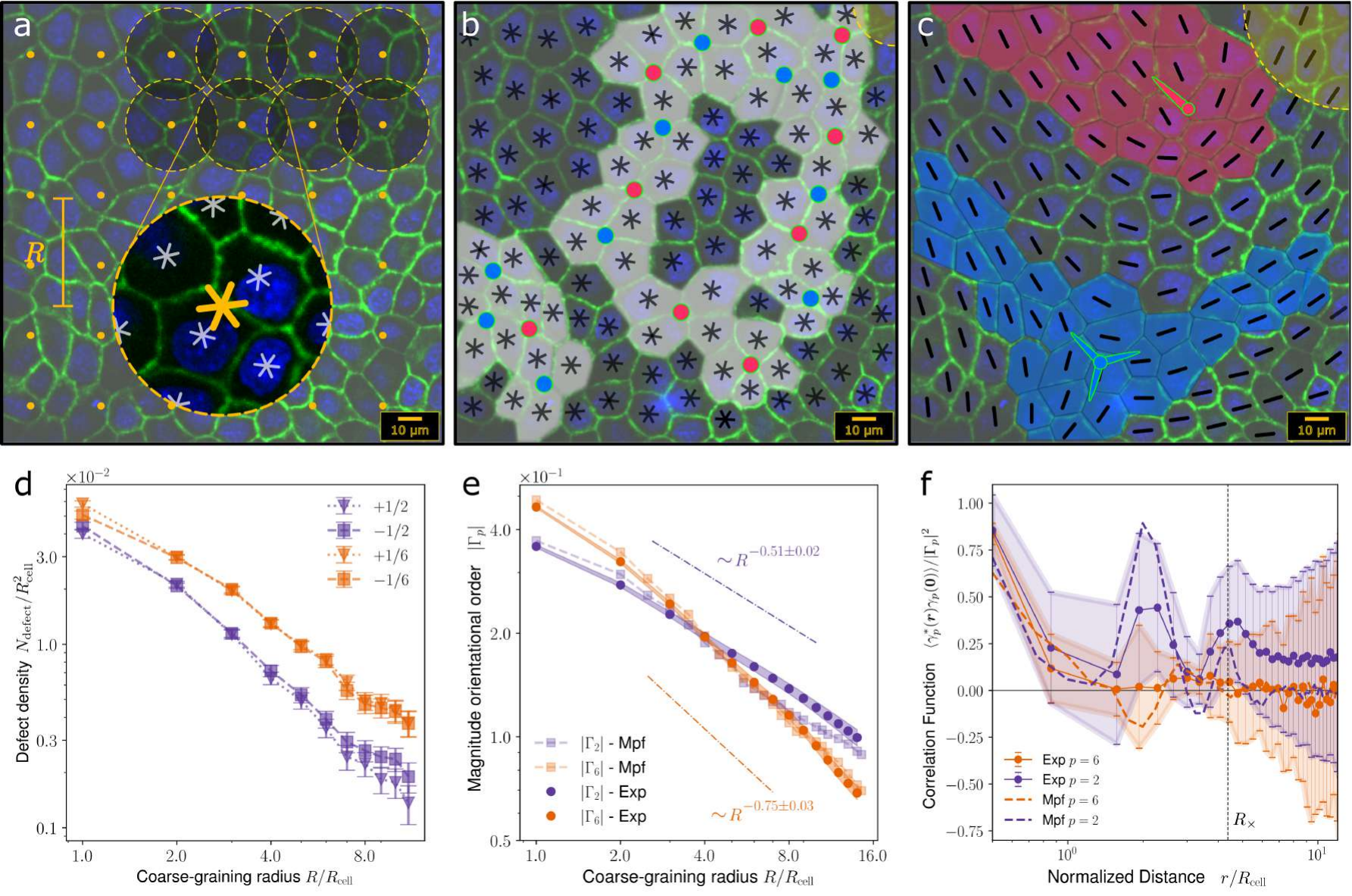}
\caption{\textbf{Coarse-graining and multiscale order in confluent epithelial monolayers.} Illustration of the coarse-graining procedure in the $p=6$ case. A lattice of (yellow) sampling points  is superimposed to the image of the cell monolayer. Each point serves as the center of a disk of radius $R$, wherein the average $\Gamma_{6}$ of the shape function $\gamma_{6}$ is computed by means of Eq.~\eqref{eqn:CoarseGraining}. The white stars mark the $6-$fold orientation of individual cells, obtained from the phase of the complex function $\gamma_{6}$, whereas the yellow star denotes the corresponding coarse-grained orientation computed from $\Gamma_{6}$. \textbf{(b)} Reconstruction of the coarse-grained $6-$fold orientation in the same configuration shown in  Fig.~\ref{Fig:fig1}a. The dashed yellow circle in the top-right corner illustrates the size of the coarse-graining domain ($R=1.5R_{\rm cell}$ in this case). Long chains of elementary hexatic defects ($+1/6$ in red, $-1/6$ in blue) separate regions characterized by uniform $6-$fold orientations. \textbf{(c)} Reconstruction of the coarse-grained $2-$fold orientation in the same configuration, but for  $R=4.5R_{\rm cell}$. Cells appears organized in chain-like structures, reminiscent of force-chains in granular materials, folding around the core of a pair of elementary nematic disclinations ($+1/2$ in red, $-1/2$ in blue). \textbf{(d)} Nematic (orange) and hexatic (purple) defect density versus the coarse-graining radius $R$ obtained from experimental data. Data are expressed in terms of the average number of defects $N_{\rm defect}$ within a square of size $R_{\rm cell}$. \textbf{(e)} $|\Gamma_{2}|$ and $|\Gamma_{6}|$ versus $R$ from experiments and numerical simulations of the multiphase-field model. Both data sets fit the power law $|\Gamma_{p}| = \langle \gamma_{p} \rangle (R/R_{\rm cell})^{-\eta_{p}/2}$. The exponents $\eta_{p}$ obtained from the fit are $\eta_{2}=1.02 \pm 0.04$ and $\eta_{6}=1.5 \pm 0.1$ (experiments); $\eta_{2}=1.30\pm 0.05$ and $\eta_{6}=1.7 \pm 0.1$ (simulations). In both experiments and simulations, $|\Gamma_{2}|$ and $|\Gamma_{6}|$ crossover at the length scale $R_{\times}$, with: $R_{\times}/R_{\rm cell}=4.3 \pm 1.0$ (experiment) and $R_{\times}/R_{\rm cell}= 4.7 \pm 1.2$ (simulations). \textbf{(f)} Correlation function $\langle \gamma_p^*(\bm{r}) \gamma_p(\bm{0})\rangle/|\Gamma_p|^2$ for experimental and numerical data.}
\label{Fig:fig3}
\end{figure*}

With the tensor $\bm{G}_{p}$ in hand, we next investigate the emergent orientational order in confluent monolayers of MDCK GII cells focussing on $2-$fold and $6-$fold symmetry (Figs.~\ref{Fig:fig1}a and~\ref{Fig:fig2}a, respectively). After segmenting the images by taking advantage of the stable expressed E-cadherin green fluorescent protein, we track the cells' contour and from the coordinates of the vertices, we compute $\gamma_{p}$ for $p=2,6$. We analyze a total of $68$ images of confluent monolayers (see Methods for details) with each one of them comprising $140 \pm 31$ cells (mean $\pm$ s.d.). Fig.~\ref{Fig:fig2}b shows the probability distribution of $|\gamma_{p}|$ for $p=2$ and $6$. Interestingly, the distribution of $|\gamma_{6}|$ is symmetric and spreads over a broad range of values; conversely, the distribution of $|\gamma_{2}|$ features a peak at approximately $0.35$, with a decreasing tail at larger values. The MDCK GII cells analyzed in this study are, therefore, more prone to arrange in hexagonal rather than elongated shapes. This results in a disordered and yet orientationally coherent tiling of the plane, where a majority of hexagons coexists with large minorities of pentagons and heptagons (see Extended Data Fig.~E3 for the distribution of the coordination number). We compare these observations with numerical simulations of two different theoretical models of epithelia: i.e. a continuous multiphase field model~\cite{Loewe2020,Monfared2021,Carenza2019} (Fig.~\ref{Fig:fig2}c) and the discrete Voronoi model~\cite{Bi2016} (Fig.~\ref{Fig:fig2}d), both in qualitative agreement with experimental data (see Methods for details of numerical models and Extended Data Fig.~E3 and E4).

In order to quantify how the shape of a cell influences the organization of other cells in its surrounding, thereby giving rise to liquid crystal order, we coarse-grain the shape function $\gamma_{p}$ over length scales larger than the typical cell radius: i.e. $R_{\rm cell}=7.4 \pm 1.9$~$\mu$m, computed as half of the distance between the centers of the segmented cells. That is 
\begin{equation}
\Gamma_{p} = \langle\gamma_{p}\rangle_{R}\;,
\label{eqn:CoarseGraining}
\end{equation}
where $\langle \dots \rangle_R$ represent the ensemble average over all cells within a disk of radius $R$ (Fig.~\ref{Fig:fig3}a and Methods). Analogously to $\gamma_{p}$, the phase $\theta=\Arg(\Gamma_{p})/p$ of this {\em shape parameter} sets the global orientation of cell clusters of size $R$, whereas its magnitude quantifies the orientational coherence of the individual cells within the cluster.

Fig.~\ref{Fig:fig3}b shows a reconstruction of the $6-$fold orientation field obtained from the computation of $\Gamma_{6}$ at a length scale only slightly larger than average cell size: i.e. $R/R_{\rm cell}=1.5$. At this scale, the monolayer appears organized in regions characterized by spatially uniform hexatic order, separated by arrays of $\pm 1/6$ disclinations, similarly to grains and grain boundaries in polycrystals~\cite{Kittel1996}. The scenario dramatically differs when the same configuration is analyzed by means of the $2-$fold shape parameter $\Gamma_{2}$, whose associated orientation field is displayed in Fig.~\ref{Fig:fig3}c for $R/R_{\rm cell}=4.5$. In this case, many of the defect structures, identified in the configuration of $\Gamma_{6}$ at the small length scales, are replaced by a smooth $2-$fold orientation filed, eventually disrupted by a few isolated $\pm 1/2$ disclinations. These, in turn, exhibit the typical comet- and star-like conformation of nematic disclinations, whereas nearby cells appear organized into chain-like structures folding around the defect cores. We stress here that, unlike previous analyses, where a $2-$fold orientation is extracted from individual cells based on their anisotropy~\cite{Duclos2016,Saw2017,Kawaguchi2017}, the rods in Fig.~\ref{Fig:fig3}c, whose orientation $\theta$ is computed from the shape parameter $\Gamma_{2}$, reflects the orientation of an entire cluster. Conversely, at smaller scales -- i.e. $R/R_{\rm cell} \lesssim 3.0$ -- the $2-$fold orientation field is characterized by very sharp and yet defect-free textures (Extended Data Fig.~E5). This peculiarity originates precisely from the mismatch between the actual $6-$fold symmetry of the configuration at the cellular scale and the $2-$fold symmetry of the order parameter used to describe it, in a similar fashion as using a (polar) vector field to describe a nematic disclination gives rise to singular lines where the polar field ``jumps'' by an angle $\pi$ (see Sec.~S2 in Supplementary Information and Extended Data Figure~E6). For both symmetries, increasing $R$ has the effect of absorbing neutral pairs of disclinations into a gently varying orientation field, resulting in a decreasing defect density (Fig.~\ref{Fig:fig3}d and Extended Data Fig.~E7 for numerical data). 

The separation of scales associated with the occurrence of nematic and hexatic features is further supported by the scaling behavior of the magnitude of $\Gamma_{2}$ and $\Gamma_{6}$ as the coarse-graining radius $R$ varies (Fig.~\ref{Fig:fig3}e). In particular, both $|\Gamma_{2}|$ and $|\Gamma_{6}|$ are finite at all length scales in the range $1 \leq R/R_{\rm cell} < 10$, but, while $|\Gamma_{6}|$ is prominent at small length scales, this is overweighted by $|\Gamma_{2}|$ at large length scales. For our MDCK GII cells on uncoated glass, the crossover occurs at $R_{\times}/R_{\rm cell}=4.3 \pm 1.0$, corresponding to clusters of order ten cells. The same crossover is also observed in our numerical simulations of the multiphase field model, with the crossover scale $R_{\times}/R_{\rm cell} = 4.7 \pm 1.2$, while it is not found in simulations of the Voronoi model, where hexatic order is dominant at all length scales. We argue that this is due to the very nature of the Voronoi model itself, which, by modeling cells as Voronoi polygons, artificially enhances their hexagonality compared to real epithelial layers, thus leading to higher hexatic order at all length scales. The multiscale structure of our MDCK layers is also reflected by the correlation function $\langle \gamma_{p}^*(\bm{r}) \gamma_{p}(\bm{0})\rangle/|\Gamma_{p}|^2$ shown in Fig.~\ref{Fig:fig3}f for experiments and multiphase-field simulations (see Methods). In particular, both nematic and hexatic correlation functions exhibit oscillations at small length scales, capturing the fine structure of cellular organization. Conversely, at larger scales, the correlations extinguish and the behavior of the correlation functions turns from oscillating to asymptotic, with the standard deviation plateauing to unity (see Extended Data Fig.~E8 for Voronoi data and time correlations). Consistently with the behavior of $|\Gamma_{p}|$, the crossover occurs at $r=R_{\times}$ (dashed vertical line). Finally, Extended Data Fig. E9 shows a plot of all $|\Gamma_{p}|$ functions for $2\le p \le 6$. Because for generic irregular polygons $\gamma_{p} \ne 0$ for all $p$ values (see e.g. Fig. S1 in Supplementary Information), the magnitude of all $\Gamma_{p}$ functions decay with $R$. Yet, $\Gamma_{2}$ and $\Gamma_{6}$ are systematically larger in magnitude than any other $\Gamma_{p}$ function, with $|\Gamma_{6}|=\max_{p}|\Gamma_{p}|$ for $R<R_{\times}$ and $|\Gamma_{2}|=\max_{p}|\Gamma_{p}|$ for $R>R_{\times}$, consistently with our interpretation.

Taken together, our experimental and numerical results demonstrate that epithelial monolayers behave as multiscale active liquid crystals, with $6-$fold hexatic order characterizing the spatial organization of the cells at small length scales, while nematic order dictates the large scale structure of the monolayer. The crossover length scale $R_{\times}$ -- i.e. the scale at which the structure of multicellular clusters changes from hexagonal to chain-like -- is, as expected, non-universal, but depends on the molecular repertoire and the material properties of the specific phenotype, as well as on the mechanical properties and the surface chemistry of the substrate~\cite{Eckert2022}. Such a chain-like structure, in turn, could result from the propagation of anisotropic stresses between neighbouring cells, in a way that is not dissimilar to the appearance of ``force chains'' in granular materials~\cite{Majmudar2005}. The possibility of a similarity between these two phenomena, has been also invoked in relation with the initial phase of Drosophila gastrulation, where linear arrays of cells simultaneously undergo apical constriction in the ventral furrow region~\cite{Gao2016}.

In conclusion, we have investigated the multiscale physics of epithelial layers finding that multiple types of liquid crystal order can coexist at different length scales. In particular, hexatic order is prominent at the cellular scale (i.e. in clusters of the order of ten cells in our MDCK GII samples) while nematic order characterizes the structure of the monolayer at larger length scales. This hierarchal structure is expected to complement the complex network of regulatory pathways that tissues have at their disposal to coordinate the activity of individual cells to achieve multicellular organization. The novel approach introduced here creates the basis for a correct identification of topological defects -- whose biophysical role in epithelia has recently focused great attention~\cite{Duclos2016,Saw2017,Kawaguchi2017}, especially in the context of morphogenesis~\cite{Keber2014,Streichan2018,Guillamat2020,Maroudas2021} -- and further provides the necessary knowledge for the foundation of a comprehensive and predictive mesoscopic theory of collective cell migration~\cite{Armengol2021}. In addition, our findings highlight a number of potentially crucial properties of epithelial tissues. First, collective cell migration in epithelia relies on both remodeling events at the small scale $-$ such as cell intercalation and the rearrangement of multicellular rosettes~\cite{Blankenship2006,Rauzi2020} $-$ as well as large scale flows~\cite{Streichan2018}. Therefore, the underlying {\em hexanematic} multiscale organization and the specific magnitude of the crossover scale $R_{\times}$ are expected to have a profound impact on how the geometry of the environment affects the specific migration strategy. E.g. metastatic cells traveling through micron-sized channels in the extracellular matrix during cancer invasion~\cite{Haeger2020} will more likely rely on local hexatic-controlled remodeling events, whereas unconfined wound healing processes~\cite{Serra2012} are more likely to leverage on system-wide nematic-driven collective flows. Second, as both hexatic and nematic liquid crystals can feature topological defects, these are expected to interact, thereby affecting processes such as the extrusion of apoptotic cells~\cite{Saw2017}, the development of sharp morphological features, such as tentacles and protrusions, in developing embryos~\cite{Maroudas2021,Hoffmann2022} and, in general, any remodelling or morphogenetic event that can take advantage of the persistent pressure variations introduced by active defects~\cite{Kawaguchi2017}. Finally, in the light of what said above, it is evident that understanding how the crossover scale $R_{\times}$ can be controlled, either chemically or mechanically, may ultimately represents the key toward deciphering tissues' collective dynamics. 

\FloatBarrier

\bibliography{Biblio.bib}

\section*{Acknowledgements}
This work is supported by the  European Union via the ERC-CoGgrant HexaTissue (L.N.C., D.K. and L.G.) and by Netherlands Organization for Scientific Research (NWO/OCW) as part of the research program ``The active matter physics of collective metastasis'' with project number Science-XL 2019.022 (J.-M.A.C and L.G.). Part of this work was carried out on the Dutch national e-infrastructure with the support of SURF through the Grant 2021.028 for computational time. J.E. and L.G. acknowledge M. Gloerich, UMC Utrecht, for providing us the MDCK cells. L.N.C. acknowledges Giuseppe Negro, University of Edinburgh, for providing us the data of passive Brownian particles. All authors acknowledge Ludwig Hoffmann for fruitful discussions.

\section*{Author Contributions}
JMAC performed analytic work, Voronoi model simulations and analyzed data. LNC coordinated the research, performed the multiphase field simulations and analyzed data. JE performed analytic work, experiments and analyzed data. DK performed analytic work and analyzed data. LG devised and coordinated the research. All authors wrote the paper. JMAC, LNC, JE and DK contributed equally to this work.

\section*{Author Information}
The authors declare no competing financial interests.

\section*{Methods}

\label{sec:mat&met}

\subsection*{Cell culture}
Parental Madin-Darby Canine Kidney (MDCK) GII cells stably expressing E-cadherin-GFP \cite{Yamada2005} (kindly provided by M. Gloerich, UMC Utrecht) were cultured in a $1:1$ ratio of low glucose DMEM (D6046; Sigma-Aldrich, St. Louis, MO) and Nutrient Mixture F-12 Ham (N4888; Sigma-Aldrich, St. Louis, MO) supplemented with $10\%$ fetal calf serum (Thermo Fisher Scientific, Waltham, MA), and $100$~mg$/$mL penicillin/streptomycin, $37$~$^\circ$C, $5\%$ CO$_2$. For experiments, cells were seeded on uncoated cover glasses, grew to confluence, and nuclei were live-stained with $2$~$\mu$g/mL Hoechst 34580 (Thermo Fisher, H21486) before imaging. 

\subsection*{Microscopy}
Samples were live-imaged with a home-build optical microscope setup based on an inverted Axiovert200 microscope body (Zeiss), a spinning disk unit (CSU-X1, Yokogawa), and an emCCD camera (iXon 897, Andor). IQ-software (Andor) was used for setup-control and data acquisition. Illumination was performed using fiber-coupling of different lasers [$405$~nm (CrystalLaser) and $488$~nm (Coherent)]. The cells on glasses were inspected with an EC Plan-NEOFLUAR $40 \times 1.3$ Oil immersion objective (Zeiss). Images were taken in three focal-planes within a distance of $352$~nm for a maximal intensity projection. The position of the nuclei was used in the image analysis as a test of single-cell segmentation.

\subsection*{Analysis}

\subsubsection*{Shape order parameter} 

Cell boundaries of confluent monolayers were analyzed using a maximum intensity projection of $z-$stack images. Cell segmentation and vertex analysis were performed using home-build Matlab scripts (Mathworks, Matlab R2018a). The number of nearest neighbors corresponds to the number of vertices surrounding a cell. The centroid of the polygon was calculated by $\bm{r}_{c}=\sum_{v=1}^{V} \bm{r}_{v} / V$ where $V$ is the number of vertices and $\bm{r}_{v}$ their positions. For each cell, the shape order was derived by using Eq.~\eqref{eqn:DefinitionOrderParameter}. On average, we analyzed $140\pm31$ cells per image. For the probability distribution of the shape order's magnitude for each analyzed image, we choose a binning of $20$ ranging from $0$ to $1$. 

\subsubsection*{Coarse-graining}  
The coarse-grained shape function $\Gamma_{p}=\Gamma_{p}(\bm{r})$, defined in Eq.~\eqref{eqn:CoarseGraining}, is constructed upon averaging the shape function $\gamma_{p}$ of the segmented cells whose center of mass, $\bm{r}_{c}$, lies within a disk of radius $R$ centered in $\bm{r}$. That is
\begin{equation}
\Gamma_{p}(\bm{r}) = \dfrac{1}{N_{\rm disk}} \sum_{c=1}^{N_{\rm cell}} \gamma_{p}(\bm{r}_c) \Theta (R - |\bm{r}-\bm{r}_c|)\;.
\label{eqn:explicit_cg}
\end{equation}
Here $\Theta(x)$ is the Heaviside step function -- such that $\Theta(x)=1$ for $x>0$ and $\Theta(x)=0$ otherwise -- and $N_{\rm disk} = \sum_{c}  \Theta (R - |\bm{r}-\bm{r}_c|)$, the number of cells whose centers lie within the disk and $N_{\rm cell}$ the total number of cells.

The position $\bm{r}$ is sampled from a square grid with lattice spacing $R_{\rm cell}$ (see Fig.~\ref{Fig:fig3}a). The radius $R$ is varied in the range $R_{\rm cell}\le R\le 88\,\mu$m, with the upper bound corresponding to half of the image size ($176\,\mu{\rm m} \times 176\,\mu{\rm m}$). Finally, the magnitude $|\Gamma_{p}|$ is averaged over different experimental/numerical configurations to obtain the curves displayed in Fig.~\ref{Fig:fig3}e (see Statistics for further details). The latter average can be accomplished in two different ways, whose difference is analogous to that between the Mean Absolute Error (MAE) and the Root Mean Square Error (RMSE) in inferential statistics. One procedure consists of averaging directly the magnitude of the shape parameter -- i.e. $\overline{|\Gamma_{p}|}$, with $\overline{(\cdots)}$ the average over different samples -- and was used in Fig.~\ref{Fig:fig3}e. Alternatively, one can average the squared amplitude -- i.e. $\overline{|\Gamma_{p}|^{2}}$ -- and then compute the square root to estimate $|\Gamma_{p}|$. Both approaches are interchangeably used in the physics of liquid crystals (see e.g. Refs~\cite{Frenkel1985,Bagchi1996}). The latter procedure gives: $\eta_{2} = 0.98 \pm 0.06$ and $\eta_{6} = 1.3 \pm 0.1$, with the crossover radius $R_{\times}=5.7 \pm 1.3$.

\subsubsection*{Topological defects} 

Topological defects were identified by first interpolating the $p-$fold orientation field on a square $22 \times 22$ grid by means of the coarse-graining procedure in Eq.~\eqref{eqn:CoarseGraining} and then computing the winding number along each unit cell (see e.g. Ref.~\cite{Giomi2015}). That is
\begin{equation}
s = \frac{1}{2\pi}\oint_{\square} {\rm d}\theta = \frac{1}{2\pi}\sum_{n=1}^{4}\left[\theta(\bm{r}_{n+1})-\theta(\bm{r}_{n})\right]\,{\rm mod}\,\frac{2\pi}{p}\;,
\end{equation}
where the symbol $\square$ denotes a square unit cell in the interpolation grid and the mod operator constraint with $\theta$ the phase of the coarse grained field $\Gamma_{p}$, as defined below Eq.~\eqref{eqn:CoarseGraining}.

\subsubsection*{Correlation Function} 
The correlation function at distance $r=|\bm{r}|$ shown in Fig.~\ref{Fig:fig3}f has been computed by averaging
\begin{multline}
\left\langle \gamma_{p}^*(\bm{r}) \gamma_{p}(\bm{0}) \right\rangle =  \frac{1}{2N_{\rm cell} N_{\rm pair}} \\ \sum_{c,c' = 1}^{N_{\rm cell}} \gamma_p^*({\bm r_c}) \gamma_p(\bm{r}_{c'}) \Pi_{r,r+\delta r}(|\bm{r}_{c}-\bm{r}_{c'}|)\;.
\end{multline}
Here $\Pi_{a,b}(x)=\Theta(x-a)-\Theta(x-b)$ is the Heaviside boxcar function -- such that $\Pi_{a,b}(x)=1$ for $a\le x \le b$ and $\Pi_{a,b}(x)=0$ otherwise -- and $N_{\rm pair}=\sum_{c,c'=1}^{N_{\rm cell}}\Pi_{r,r+\delta r}(|\bm{r}_{c}-\bm{r}_{c'}|)$ the number of pairs of cells whose centers are at a distance $r \le |{\bm r_c} - {\bm r_c'} | \le r+\delta r$ from each other. To guarantee a sufficient spatial resolution, the the binning distance $\delta r$ must be chosen to be smaller than the typical size of cells. In Fig.~\ref{Fig:fig3}, $\delta r/R_{\rm cell} = 0.4$.

\subsection*{Statistics}

In total, 68 images of confluent monolayers (nine coverslips, three independent experiments) were taken and analyzed. In total, 9496 cells were considered for the analysis.

\subsection*{Numerical simulations}
We make use of two different numerical models for ET previously introduced in literature: \emph{(i)} the multiphase field model and \emph{(ii)} the Voronoi model. 

\subsubsection*{Multiphase field model} 

This model has been used to study the dynamics of confluent cell monolayers~\cite{Loewe2020} and the mechanics of cell extrusion~\cite{Monfared2021}. It is a continuous model where each cell is described by a concentration field $\varphi_{c}=\varphi_{c}(\bm{r})$, with $c=1,\,2\ldots\,N_{\rm cell}$. The equilibrium state is defined by the free energy $\mathcal{F}=\int {\rm d}A\,f $ where the free energy density $f$ is given by
\begin{multline}\label{eqn:freeenergy_multiphase}
f 
= \frac{\alpha}{4} \sum_{c} \varphi_{c}^{2}(\varphi_{c}-\varphi_{0})^{2} 
+ \frac{k_{\varphi}}{2}\sum_{c}(\nabla\varphi_{c})^{2} \\ 
+ \epsilon\sum_{c<c'}\varphi_{c}^{2}\varphi_{c'}^{2} 
+ \sum_{c} \lambda\left(1-\frac{1}{\pi \varphi_{0}^{2}R_{\varphi}^{2}} \int {\rm d}A\,\varphi_{c}^{2}\right)^{2}\;.
\end{multline}
Here $\alpha$ and $k_\varphi$ are material parameters which can be used to tune the surface tension $\sigma = \sqrt{8\kappa_\varphi \alpha}$ and the interfacial thickness $\xi=\sqrt{2 \kappa_\varphi /\alpha}$ of isolated cells and thermodynamically favor spherical cell shapes. The constant $\epsilon$ captures the repulsion between cells. The concentration field is large (i.e. $\varphi_{i} \simeq \varphi_{0}$) inside the cells and zero outside. The contribution proportional to $\lambda$ in the free energy enforces cell incompressibility whose nominal radius is given by $R_{\varphi}$.
The phase field $\varphi_i$ evolves according to the Allen-Cahn equation
\begin{equation}\label{eqn:allen_cahn}
\partial_{t} \varphi_{c}+\bm{v}_{c} \cdot \nabla \varphi_{c} = - M\,\frac{\delta \mathcal{F}}{\delta \varphi_{c}}\;,
\end{equation}
where $\bm{v}_{i}=v_{0}(\cos\theta_{c}\,\bm{e}_{x}+\sin\theta_{c}\,\bm{e}_{y})$ is the velocity at which the $c-$th cell self-propels, with $v_{0}$ a constant speed and $\theta_{c}$ an angle. The latter evolves according to the stochastic equation
\begin{equation}
\frac{{\rm d}\theta_{c}}{{\rm d}t} = \eta_{c}\;,
\end{equation}
where $\eta_{c}$ is a noise term with correlation function $\langle \eta_{c}(t)\eta_{c'}(t') \rangle = 2D_{\rm r}\delta_{cc'}\delta(t-t')$ and $D_{r}$ a constant controlling noise diffusivity. The constant $M$ in Eq.~\eqref{eqn:allen_cahn} is the mobility measuring the relevance of thermodynamic relaxation with respect to non-equlibrium cell migration. Eq.~\eqref{eqn:allen_cahn} is solved with a finite-difference approach through a predictor-corrector finite difference Euler scheme implementing second order stencil for space derivatives~\cite{Carenza2019}. Simulation details and scaling to physical units are given in Table~\ref{Table:simulation_parameters}.

\subsubsection*{Voronoi model}

This model portrays a confluent tissue as a Voronoi tesselation of the plane~\cite{Bi2016}. Each cell is characterized by two dynamical variables: the position $\bm{r}_{c}$ and the velocity $\bm{v}_{c}=v_{0}(\cos\theta_{c}\,\bm{e}_{x}+\sin\theta_{c}\,\bm{e}_{y})$ with $v_{0}$ a constant speed and $\theta_{c}$ an angle, with $c=1,\,2\ldots N_{\rm cell}$. The dynamics of these variables is governed by the following set of ordinary differential equations
\begin{subequations}\label{eqn:voronoi_eom}
\begin{gather}
\frac{{\rm d}\bm{r}_{c}}{{\rm d}t} = \bm{v}_{c}-\mu\nabla_{\bm{r}_{c}}E\;,\\
\frac{{\rm d}\theta_{c}}{{\rm d}t} = \eta_{c}\;,
\end{gather}
\end{subequations}
where $\mu$ is a mobility coefficient and $E=E(\bm{r}_{1},\bm{r}_{2}\ldots\,\bm{r}_{N_{\rm cell}})$ is an energy function defined as
\begin{equation}\label{eqn:voronoi_energy}
E = \sum_{c}\left[K_{A}\left(A_{c}-A_{0}\right)^{2}+K_{P}\left(P_{c}-P_{0}\right)^{2}\right]\;.
\end{equation}
Here $A_{c}$ and $P_{c}$ are respectively the area and perimeter of each cell and $A_{0}$ and $P_{0}$ their preferred values. The variable $\eta_{c}$ in Eq.~(\ref{eqn:voronoi_eom}b) is white noise, having zero mean and correlation function
\begin{equation}
\langle \eta_{c}(t)\eta_{c'}(t') \rangle = 2D_{\rm r}\delta_{cc'}\delta(t-t')\;,
\end{equation}
with $D_{\rm r}$ a rotational diffusion coefficient. Simulation details and scaling to physical units are given in Table~\ref{Table:simulation_parameters}.

\subsubsection*{Dimensionless numbers}

Two dimensionless numbers can be extracted from both the multiphase field and Voronoi models.
These are

\begin{itemize}
	
\item The  active P{\'e}clet number $\rm Pe$, measuring the relevance of activity with respect to noise diffusivity. For the multiphase field model ${\rm Pe}=v_{0}R_{\varphi}/D$; for the Voronoi model ${\rm Pe}=v_{0}\sqrt{A_0}/D$;

\item The cell deformability. In the multiphase field model this is given by $d=\epsilon/\alpha$, namely the ratio of cell-cell repulsion and bulk compressibility. For the Voronoi model cell compressibility is expressed in terms of the shape factor $p_0= P_0/\sqrt{A_{0}}$.

\item The total cell density.

\end{itemize}

The parameters values have been specifically chosen to maximize the agreement with the experimental results.
More in detail, multiphase field simulations simulations were performed with cell deformability $d=1.66$. This is to ensure that for large enough total density, the model cells attain a polygonal shape and settle into a confluent configuration at steady state, see for instance Ref.~\cite{Loewe2020} [notice that equilibrium shape is circular for isolated cells as prescribed by the field theory in Eq.~\eqref{eqn:freeenergy_multiphase}].
Voronoi model simulations were performed with $p_0=3.9$ corresponding to an liquid system in the passive limit ($v_0=0$).  
The P{\'e}clet number $\rm Pe$ was varied in the range from $0$ to $5.0$ for multiphase field simulations and from $0$ to $0.25$ in Voronoi simulations. The results presented in the main text refer to the case ${\rm Pe}=3.22$ and ${\rm Pe}=0.1$, for multiphase field and Voronoi model, respectively.
These values were chosen to minimize the $L_{2}$-distance between numerical and experimental $|\gamma_6|$ distributions. 
Refer to Table~\ref{Table:simulation_parameters} for a complete list of parameters.

\begin{table*}[b]
\centering
\caption{\textbf{Physical scaling of simulation parameters.}
The table provides the parameters used to perform simulations for both the multiphase field and the Voronoi  model, together with their dimensions and scaling to physical units. For the multiphase field model, scaling is performed by equating the mean cell radius $R_{\rm cell}$ ($\simeq 7.4 \mu \text{m}$) measured in experiments with the nominal cell radius $R$ and a typical migration speed of cells in MDCK monolayers~\cite{Balasubramaniam2021} ($\simeq 2 \mu \text{m} \ \text{h}^{-1}$) with that measured in our simulations ($\simeq 0.0011 \Delta x/ \Delta t$). This allows us to find the physical scaling of the lattice grid unit $\Delta x$ and the iteration unit $\Delta t$.
For the Voronoi model, we equated the mean cell radius $R_{\rm cell}$ in experiments with that measured in simulations ($\simeq 1$). The time-step was derived with the same procedure as described for the multiphase field model.
 In the table, simulation values are given in both lattice and physical units, in columns four and five, respectively. Notice that we did not introduce an energy scale as this cancels out with the mobility parameter $M$ in Eq.~\eqref{eqn:allen_cahn} and $\mu$ in Eq.~\eqref{eqn:voronoi_eom}, respectively.
}
\begin{tabular}{l!{\vrule width \heavyrulewidth}llll} 
\toprule
\multicolumn{5}{c}{\textbf{Numerical model}}                                                                                                                                                                                              \\ 
\toprule
\multicolumn{5}{c}{\textit{Multiphase field model}}                                                                                                                                                                                       \\ 
\toprule
\multicolumn{2}{c}{Model parameter}                                                                                     & \multicolumn{1}{c}{Dimension} & \multicolumn{1}{c}{Simulation value(s)} & \multicolumn{1}{c}{Physical scaling}  \\ 
\toprule
\multirow{4}{*}{\textit{Lattice parameters}}           & $N_{\rm cell}$                                                       & ---          				& $361$                                   & ---            \\
                                                       & $\Delta x$                                                     & $L$                           & $1$                                     & $0.685 \ \mu $m                                \\
                                                       & $\Delta t$                                                     & $T$                           & $1$                                     & $1.414 $ s                                      \\
                                                       & $L_x,L_y$                                                      & $L$                           & $380$                                   &  $246.6 \ \mu $m              		 \\ 
\cmidrule{1-2}\cmidrule[\heavyrulewidth]{3-5}
\multirow{8}{*}{\textit{Free energy parameters}}       & $M \alpha$                                                     & $1/T$                         & $0.006$                                 &  $0.0042 \ \text{s}^{-1}$                                     \\
                                                       & $M k_\varphi$                                                  & $L^2/T$                       & $0.012$                                 &  $0.0040 \ \mu \text{m}^2 \ \text{s}^{-1}$              \\
                                                       & $M \epsilon$                                                   & $1/T$                         & $0.01$                                  & $0.0071 \ \text{s}^{-1}$                                     \\
                                                       & $M \lambda$                                                    & $1/T$                         & $600$                                   & $424,4 \ \text{s}^{-1}$                                     \\
                                                       & $\varphi_0$                                                    & ---         					& $2.0$                                   & ---				  	                  \\
                                                       & $R$                                                            & $L$                           & $10.86$                                 & $7.4 \ \mu\text{m}$                                     \\
                                                       & $\xi=\sqrt{2 k_\varphi/\alpha}$                                & $L$                           & $2$                                     & $1.37 \ \mu\text{m}$                                     \\
                                                       & $M\sigma = M\sqrt{8/9 k_\varphi \alpha}$                       & $L/T$                         & $0.008$                                 & $0.0039 \ \mu \text{m s}^{-1}$                                     \\
\cmidrule{1-1}\cmidrule[\heavyrulewidth]{2-5}
\multirow{2}{*}{\textit{Dynamical equation parameters}} & $D^{pf}_r$                                                          & $1/T$                         & $0.0001$                                & $0.00007 \ \text{s}^{-1}$                                     \\
                                                       & $v_0$                                                          & $L/T$                         & $0.0035$                                & $0.00169 \ \mu \text{m s}^{-1}$                                     \\
\cmidrule{1-1}\cmidrule[\heavyrulewidth]{2-5}
\multirow{2}{*}{\textit{Dimensionless numbers}}        & Pecl{\'e}t number ${\rm Pe}=v_0/(D_r R)$             & ---				            & 3.22          & ---					                  \\
                                                       & Cell deformability $d=\epsilon/\alpha$                                         & ---				            &                                         1.66 & ---				                	  \\ 
\toprule
\multicolumn{5}{c}{\textit{Voronoi model}}                                                                                                                                                                                                \\ 
\toprule
\multirow{3}{*}{\textit{Lattice parameters}}           & $N_{\rm cell}$                                                     & ---				            & $22500$                                   & ---				            \\
                                                       & $\Delta t$                                                     & $T$                           &  $0.01$                                       &$0.53\ \text{s}$                                      \\
                                                       & $L_x,L_y$                                                      & $L$                           &    $150$                                     &   $2220 \mu\text{m}$                                    \\ 
\toprule
\multirow{4}{*}{\textit{Energy parameters}}            & $\mu K_A$                                                      & $1/(L^2T)$                    &    $1$                                     &   $0.0086  \ \mu\text{s}^{-1}$                                    \\
                                                       & $\mu K_P$                                                      & $1/T$                         &   $1$                                      &  $0.019\ \text{s}^{-1}$                                     \\
                                                       & $A_0$                                                          & $L^2$                         &         $1$                                &     $219.04\ \mu \text{m}^{-2}$                                  \\
                                                       & $P_0$                                                          & $L$                           &  $3.9$                                       &    $57.72\ \mu \text{m}$                                   \\ 
\toprule
\multirow{2}{*}{\textit{Dynamical equation parameters}} & $v_0$                                                          & $L/T$                         &    $0.1$                                     &             $0.00278 \ \mu \text{m} \ \text{s}^{-1}$                          \\
                                                       & $D'_r$                                                       & $1/T$                         &     $1$                                    &     $0.019\ \text{s}^{-1}$                                  \\ 
\toprule
\multirow{2}{*}{\textit{Dimensionless numbers}}        & Pecl{\'e}t ${\rm Pe}=v_0/(D^{V}_r \sqrt{A_0})$ & ---				            &            $0.1$                             & ---				            \\
                                                       & Shape index $p_0=P_0/\sqrt{A_0}$                               &  ---				            &  $3.9$                                       & ---				            \\
\toprule
\end{tabular}
\label{Table:simulation_parameters}
\end{table*}

\end{document}


\title{Supplementary information}

\maketitle

\tableofcontents

\section{\label{sec:1}Orientational order in two-dimensional $p-$atic liquid crystals}

\subsection{\label{sec:1a}Definition and basic properties}

In this supplementary section we review some basic concepts of orientational order in two-dimensional liquid crystals. To this end, let us consider a planar arrangement of $N$ {\em regular} and equally sized $p-$sided polygons (not necesserily confluent), whose position and orientation are denoted by $\bm{r}=x\,\bm{e}_{x}+y\,\bm{e}_{y}$ and  $\bm{\nu}=\cos\vartheta\,\bm{e}_{x}+\sin\vartheta\,\bm{e}_{y}$, with $\{\bm{e}_{x},\bm{e}_{y}\}$ orthonormal basis vectors. The rotational symmetry of the polygons introduces some arbitrariness in the choice of the angle $\vartheta$ (see Fig.~\ref{fig:S1}a), which, nonetheless, can be removed upon constructing a $p-$fold symmetric order parameter. The latter is obtained by introducing the complex function
\begin{equation}\label{eq:psi}
\psi_{p} = e^{ip\vartheta}\;,
\end{equation}
which is manifestly invariant under rotation by $2\pi/p$, thus compensating for the arbitrariness in the choice of the angle $\vartheta$. The function $\psi_{p}$ can then be averaged over the length scale $R$ to obtain the $p-$atic order parameter
\begin{equation}\label{eq:Psi}
\Psi_{p} = \langle \psi_{p} \rangle_{R}\;.
\end{equation}
The scale-average $\langle\cdots\rangle_{R}$ can be computed in different ways depending on the setting of the problem. In equilibrium and non-equilibrium field theories
\begin{equation}
\langle \cdots \rangle_{R}=\int\prod_{q_{\min}<|\bm{q}|<q_{\max}}{\rm d}\vartheta_{\bm{q}}\,P(\vartheta)(\cdots)\;, 
\end{equation}
with $\vartheta_{\bm{q}}$ is the Fourier amplitude associated with the wave vector $\bm{q}$, $P=P(\vartheta)$ the probability distribution of the filed $\vartheta$ and $q_{\min}=2\pi/R$ and $q_{\max}=2\pi/a$, with $a$ a short distance cut-off, the lower (i.e. infrared) and upper (i.e. ultraviolet) bound of the wavenumber $|\bm{q}|$ in the range of length scales going from $a$ to $R$. In experiments and numerical simulations, on the other hand, $\langle \cdots\rangle_{R}$ can be computed by averaging over all the polygons within a distance $R$ from a given reference position (see Methods). Because of the lack of long-ranged order in two-dimensional liquid crystals, its magnitude decays as a power law for increasing $R$ values: i.e.
\begin{equation}\label{eq:power_law_scaling}
\left|\Psi_{p}\right| \sim R^{-\eta_{p}/2}\;,
\end{equation}  
and $\eta_{p}$ a non-universal exponent expressing the rate of power law decay of the orientational correlation function
\begin{equation}
\label{eq:psi_correlation}
\left\langle\psi_{p}^{*}(\bm{r})\psi_{p}(\bm{0})\right\rangle \sim |\bm{r}|^{-\eta_{p}}\;.
\end{equation}
At equilibrium, $0<\eta_{p}\le 1/4$, with the upper bound corresponding to the Kosterlitz-Thouless transition, where orientational order is lost due to disclination unbinding (see e.g. Ref.$^{\href{https://doi.org/10.1103/PhysRevLett.41.121}{23},\,\href{https://doi.org/10.1103/PhysRevB.19.2457}{24}}$). In active liquid crystals, where defect unbinding is caused by internal stresses$^{\href{https://doi.org/10.1103/PhysRevX.5.031003}{44}}$, such an upper bound is expected to be larger and possibly reach the value $\eta_{p}=2$, corresponding to a uniformly distributed orientation field.

\subsection{\label{sec:1b}Regular versus irregular polygons}

\begin{figure*}[t!]
\centering
\includegraphics[width=\textwidth]{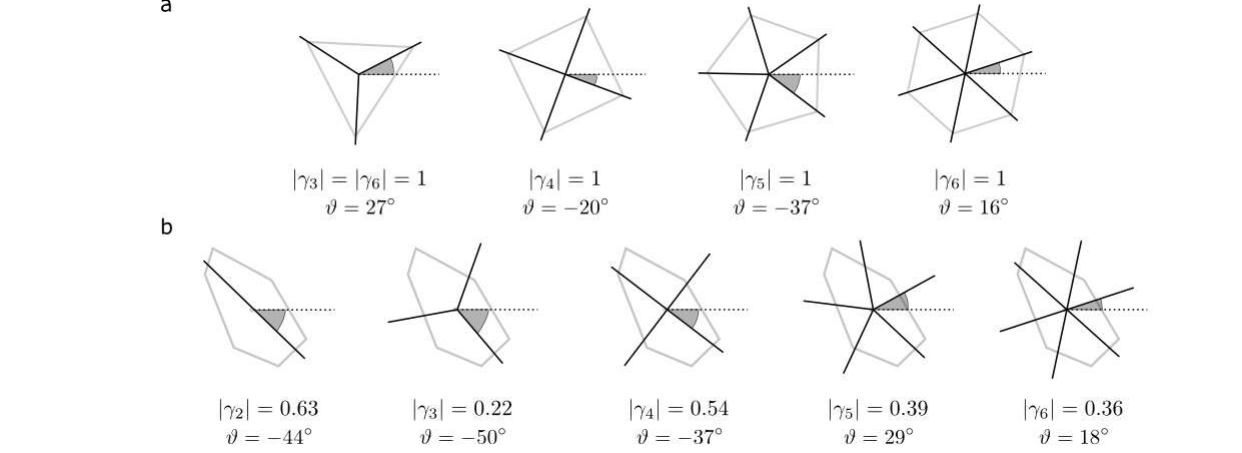}
\caption{\label{fig:S1}\textbf{Shape function $\gamma_{p}$.} Example of the shape function $\gamma_{p}$ for polygons with different degree of regularity. The black star at the center of each polygon marks its orientation calculated via the phase $\vartheta=\Arg(\gamma_{p})/p$, which is, in turn, highlighted in grey. (a) In the case of regular $V-$sided polygons, $|\gamma_{p}|=\delta_{nV,p}$, with $n\in\mathbb{N}$. (b) For irregular polygons, on the other hand, the magnitude of $\gamma_{p}$ quantifies the resemblance between the polygon and a regular $p-$sided polygon having the same size. Because of its elongated shape, the $6-$sided polygon displayed here has a prominent $2-$fold structure, consistent with the fact that $\gamma_{2}$ is larger in magnitude than any of the other shape functions.}
\end{figure*}

The construction reviewed in Sec.~\ref{sec:1a} demands assigning an orientation $\vartheta$ to each individual building block of the system. For segments (i.e. $p=2$) and regular polygons (i.e. $p\ge 3$), the latter is readily achieved by defining $\vartheta$ as the orientation, relative to the $x-$axis of a standard Cartesian frame, of the closest vertex (see Fig.~\ref{fig:S1}a). As previously explained, the invariance of the function $\psi_{p}$ under transformations of the form $\vartheta \to \vartheta+2\pi k/p$, with $k\in\mathbb{Z}$, accounts for the rotational symmetry of the building blocks. In the case of irregular polygons, on the other hand, the same criterion cannot be used, because of the lack of an exact rotational symmetry. In order to overcome this limitation and define a $p-$fold orientation of an arbitrarily shaped $V-$sided polygon we have introduced in the main text the shape function
\begin{equation}\label{eq:gamma}
\gamma_{p} = \frac{1}{\Delta_{p}} \sum_{v=1}^{V}|\bm{r}_{v}|^{p}e^{ip\phi_{v}}\;,
\end{equation}
where $\bm{r}_{v}$ and $\phi_{v}$ are respectively the position and orientation of the $v-$th vertex of the polygon with respect to its center of mass and $\Delta_{p}=\sum_{v=1}^{V}|\bm{r}_{v}|^{p}$ is a normalization constant. In Sec.~\ref{sec:2a}, we demonstrate how the real and imaginary parts of this  complex function corresponds to the only independent components of the rank$-p$ tensor $\bm{G}_{p}$, given in Eq.~(1) of the main text.

Compared to the classic $p-$atic order parameter, the function $\gamma_{p}$ has the twofold advantage of capturing both the degree of regularity of the polygon -- through its amplitude $0\le |\gamma_{p}| \le 1$ -- as well as its orientation -- via its phase. To illustrate these concepts, it is particularly useful to consider the special case of regular polygons. In this case, $\gamma_{p}$ only contains information on the orientation (i.e. $|\gamma_{p}|=1$) and reduces to the $p-$atic order parameter $\psi_{p}$. To demonstrate the latter statement, we notice that, in a regular $V-$sided polygon, $\phi_{v}=\vartheta+2\pi (v-1)/V$, with $v=1,\,2\ldots V$ and $\vartheta$ the angular position of the vertex closer to the $x-$axis. In this case, $|\bm{r}_{v}|$ is the same for all the vertices and equates the circumradius of the polygon for all $v$ values. From Eq.~\eqref{eq:gamma} it follows that
\begin{equation}
\gamma_{p} 
= \frac{e^{ip\vartheta}}{V}\sum_{v=1}^{V}e^{2\pi i p\,\frac{v-1}{V}}	
= e^{ip\vartheta}\delta_{nV,p}\;,\qquad n\in\mathbb{N}\;.
\end{equation} 
Thus, if the integer $p$ is a multiple of $V$, $\gamma_{p}=\psi_{p}$. This algebraic property reflects the orientational degeneracy of regular $V-$sided polygons, which are invariant under rotations by any multiple of $2\pi/V$: e.g. a regular square is symmetric under rotations by $\pi/2$, $\pi$, $3\pi/2$, $2\pi$ etc. Conversely, $\gamma_{p}=0$ for $V > p$. A simple but instructive example of the latter scenario can be obtained in the case $p=2$ and $V=4$, for which computing the phase of the shape function $\gamma_{2}$ is equivalent to tracing the diagonal of a square. The latter can be oriented at four different angles (e.g. the same angles as in the previous example), corresponding to two pairs of opposite directions in the complex plane, so that $\gamma_{2}=0$. The same argument holds for arbitrary $V=2p$ values.

For irregular polygons, on the other hand, the complex function $\gamma_{p}$ is in general finite and larger in magnitude for shapes that better approximate a regular $p-$sided polygon (see Fig.~\ref{fig:S1}b for an illustration and Sec.~\ref{sec:g_tensor} for a proof of the latter statement). Intuitively, this results from the $|\bm{r}_{v}|^{p}$ term in Eq.~\eqref{eq:gamma}, which assigns a larger weight to the vertices that are further from the center of the polygon, thus have a larger impact on the polygon overall shape. As for the $p-$atic order parameter, one can define a {\em shape parameter} upon averaging over the length scale $R$. That is
\begin{equation}\label{eq:Gamma}
\Gamma_{p} = \langle\gamma_{p}\rangle_{R}\;.
\end{equation}
Now, whether regular or irregular, the local $p-$fold orientation of a polygon can be defined, in analogy with the construction reviewed in Sec.~\ref{sec:1a} from the phase of the function $\gamma_{p}$. That is
\begin{equation}\label{eq:vartheta}
\vartheta = \frac{\Arg(\gamma_{p})}{p} = \frac{1}{p}\,\arctan \left[\frac{\Im(\gamma_{p})}{\Re(\gamma_{p})}\right]\;,
\end{equation}
where $\Re$ and $\Im$ denotes the real and imaginary part, respectively. Finally, using Eqs.~\eqref{eq:psi} and \eqref{eq:vartheta}, we obtain the general relation
\begin{equation}
\psi_{p} = \frac{\gamma_{p}}{|\gamma_{p}|}\;,	
\label{eq:psip_and_gamma}
\end{equation}
from which one can generalize the notion of orientational order to arbitrary polygons, thereby introducing what will be hereafter referred to as {\em shape} orientational order. We stress that the shape parameter $\Gamma_p$ inherits the scaling behavior of the order parameter $\Psi_{p}$ so that the correlation function of the shape-orientational order $\left\langle\gamma_{p}^{*}(\bm{r})\gamma_{p}(\bm{0})\right\rangle \sim |\bm{r}|^{-\eta_{p}}$ scales with the same exponent $\eta_{p}$ as in Eq.~\eqref{eq:psi_correlation}. These statement will be proved in Sec.~\ref{sec:1d} where we compare the parameters $\Gamma_{p}$ ad $\Psi_{p}$, as well as the correlation functions $\langle\gamma_{p}^{*}(\bm{r})\gamma_{p}(\bm{0})\rangle$ and  $\langle\psi_{p}^{*}(\bm{r})\psi_{p}(\bm{0})\rangle$, for the case of a two-dimensioanl system of passive Brownian particles on the verge of melting -- a system which is well-known to exhibit hexatic order and that we use here to benchmark our parameter for shape orientational order.

\subsection{\label{sec:1c}Shape versus bond orientational order}

\begin{figure*}[t!]
\centering
\includegraphics[width=\textwidth]{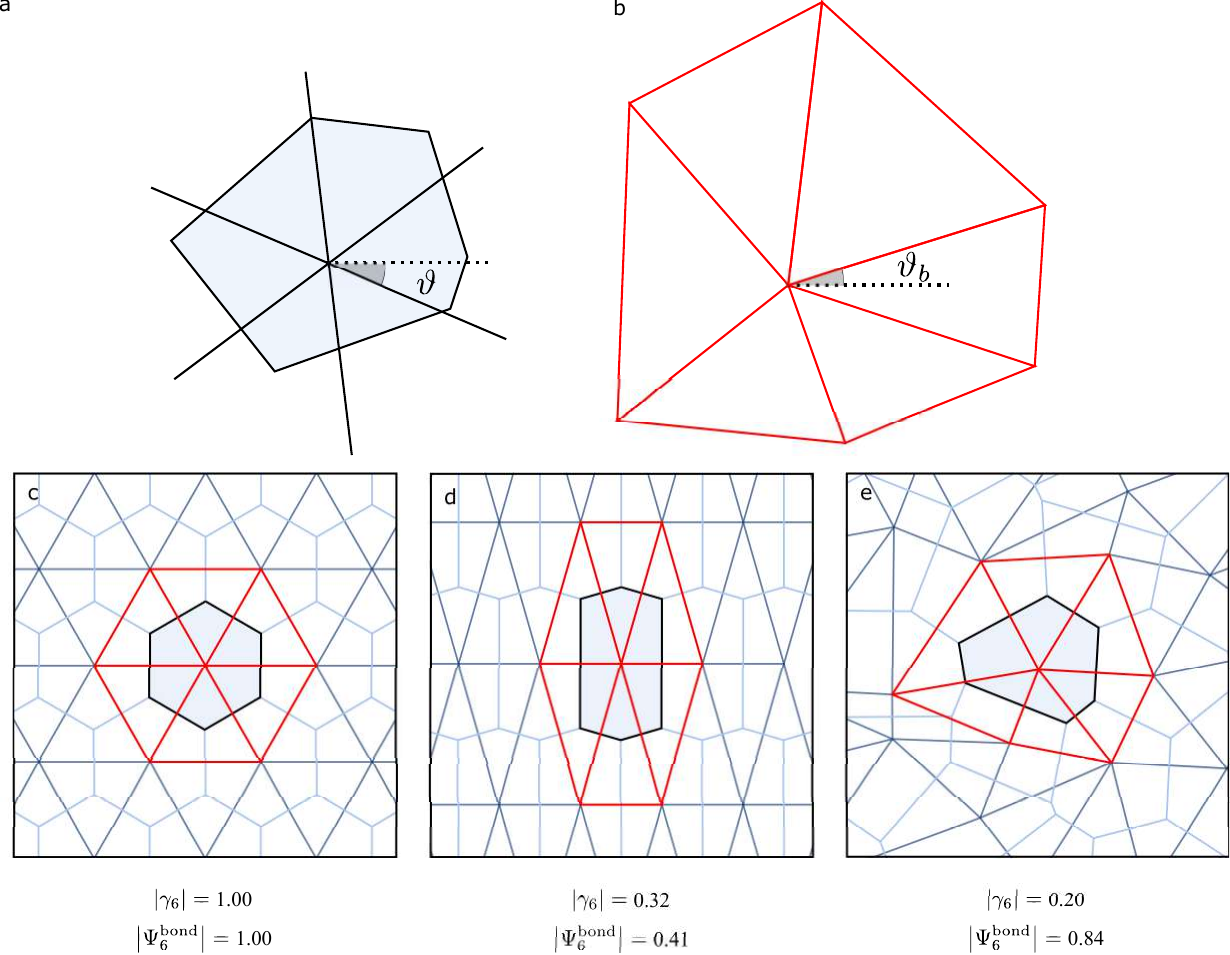}
\caption{\label{fig:S2}\textbf{Shape versus bond order.} (a) A Voronoi cell with its $6-$fold orientation $\vartheta$ computed from the shape function $\gamma_{6}$. (b) Delaunay graph associated with the Voronoi cell, with $\vartheta_{b}$ the orientation of the bond closest to the $x-$axis. The black dot labeled as $D$ in (a) marks the position of the central Delaunay vertex, whereas that labelled as $V$ in (b) corresponds to the centroid of the Voronoi cell. (c-e) Three examples of tesselations with different degree of regularity and corresponding values of $|\gamma_{6}|$ -- i.e. Eq.~\eqref{eq:gamma} -- and $|\Psi_{6}^{\rm bond}|$ -- i.e. Eq.~\eqref{eq:Psi_bond}.}
\end{figure*}

Although the concepts summarized in the previous two sections naturally apply to assemblies of polygons, whose microscopic structure is inherited by the system at larger length scales, $p-$atic order can be also found in collections of isotropic particles or, generally, in any planar network consisting of nodes connected by bonds. For the special case of hexatics, such a $6-$fold {\em bond} orientational order was first unveiled in the context of the so called Kosterlitz-Thouless-Halperin-Nelson-Young (KTHNY) melting scenario$^{\href{https://doi.org/10.1103/PhysRevLett.41.121}{23},\,\href{https://doi.org/10.1103/PhysRevB.19.2457}{24}}$ and can be readily generalized to generic $p-$fold symmetry via Eq.~\eqref{eq:psi}, by replacing the orientation $\vartheta$ of a segment (i.e. $p=2$) or regular polygon (i.e. $p\ge 3$ and Fig.~\ref{fig:S1}a), with the orientation $\vartheta_{b}$ of {\em any} of the bonds emanating from a node of a given bond network (see Fig.~\ref{fig:S2}b). To distinguish this function from that defined in Eq.~\eqref{eq:psi}, we introduce, in this and the next Section only, the notation 
\begin{equation}
\psi_{b}^{\rm bond}=e^{ip\vartheta_{b}}\;,
\end{equation}
from which, using Eq.~\eqref{eq:Psi}, one can construct the coarse-grained order parameter $\Psi_{p}^{\rm bond}=\langle \psi_{p}^{\rm bond}\rangle_{R}$. When the coarse-graining scale $R$ equates the typical size of a {\em local neighborhood}, in particular, this quantity takes the form
\begin{equation}\label{eq:Psi_bond}
\Psi_{p}^{\rm bond} = \frac{1}{B}\sum_{b=1}^{B} e^{ip\vartheta_{b}}\;,	
\end{equation}	
with $B$ the total number of bonds emanating from the central site of the neighborhood (see Fig.~\ref{fig:S2}c-e). Eq.~\eqref{eq:Psi_bond} is the most common definition of $p-$atic bond orientational order parameter in numerical studies$^{\href{https://doi.org/10.1103/PhysRevLett.107.155704}{25},\,\href{https://doi.org/10.1103/PhysRevLett.121.098003}{26}}$. 

Now, because it is always possible to construct a bond network from a generic planar tessellation and vice versa, it is natural to ask whether the shape function $\gamma_{p}$ and the bond order parameter $\Psi_{p}^{\rm bond}$ contain the same information. To answer this question we have reported in Figs.~\ref{fig:S2}c-e three examples of planar tessellations with different degrees of regularity. In all panels, polygons consist of Voronoi cells constructed from a preexisting set of control points, whereas the bond network is obtained from the associated Delaunay triangulation. With only exception for Fig.~\ref{fig:S2}c, corresponding to a regular honeycomb lattice, $\gamma_{p}$ is always different from $\Psi_{p}^{\rm bond}$. 

We also stress that there is no unique bond network that one could associate to a given planar tessellation and, vice versa, there is no unique planar tessellation that one could construct starting from a given bond network. For instance, starting from the same bond network shown in Figs.~\ref{fig:S2}c-e one could construct an alternative Voronoi tessellation in which the generating point of each Voronoi cell is also its centroid (center of mass). Analogously, given a planar tessellation, constructing an associated bond network demands assigning a center to each cell, hence a criterion that, without additional insight into the physical mechanisms involved in the system, is also arbitrary. From these considerations, we conclude that the most physically meaningful orientational characterization of confluent cell layers, foams or other cellular structures, can be accomplished via the shape function $\gamma_{p}$, whereas for assemblies of isotropic particles, such as the original hexatic phase introduced by Halperin and Nelson$^{\href{https://doi.org/10.1103/PhysRevLett.41.121}{23},\,\href{https://doi.org/10.1103/PhysRevB.19.2457}{24}}$, via $\psi_{p}^{\rm bond}$. 

\subsection{\label{sec:1d} A benchmark analysis: Orientational order in the melting transition of passive Brownian particles }

\begin{figure}[t!]
\includegraphics[width=\textwidth]{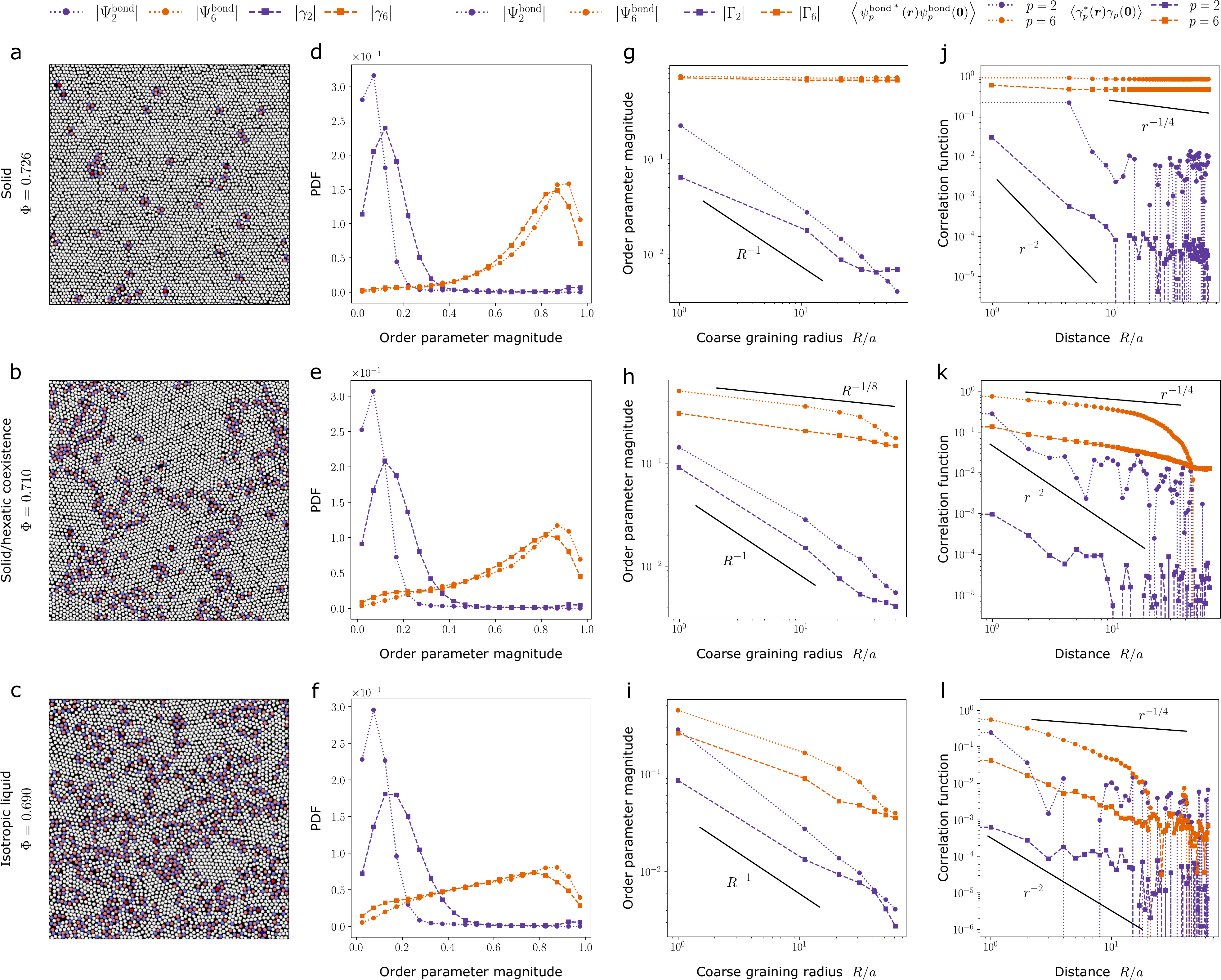}
\caption{\label{fig:S3}\textbf{Orientational order in systems of passive Brownian particles.} Comparison between bond and shape orientational order in a system of passive Brownian particles across the solid-liquid transiton for three different values of the packing fraction $\Phi$: i.e. $0.726$ (first row, solid phase), $0.710$ (second row, solid/hexatic phase coexistence) and $0.690$ (third row, isotropic liquid phase). (a-c) typical configuration of a subset of the entire system after thermalization. Particles with $5$ ($7$) nearest neighbors are highlighted in red (blue), while other particles are colored in white. (d-f) The probability distribution of the magnitude of the complex function $\gamma_{p}$ and $\Psi_{p}^{\rm bond}$ for $p=2,\,6$. (g-i) Same quantities for increasing length scale $R$. Notice that, irrespectively of the phase of the system (solid, liquid or in between), both order parameters decays with the same power-law $\eta_{p}/2$. (j-l) Two-point correlation functions of $\gamma_{p}$ and $\psi_{p}$ for $p=2,\,6$. As expected, the scaling behavior is consistent with that observed in the coarse-grained shape function and bond orientional order parameter.}
\end{figure}

In this Section we will benchmarck the different parameters introduced so far to capture orientational order on a standard passive system consisting of a two-dimensional collection of passive Brownian particles (PBP) on the verge of melting, a process whose characterization in terms of orientational order is established. The particles interact via a hard-core repulsions and are subject to thermal fluctuations. At low packing fractions -- i.e. $\Phi \lesssim 0.700$ -- the system is in an isotropic liquid phase, with no positional nor orientational order. For $\Phi>0.720$, on the other hand, a continuous hexatic-solid transition takes place, in accordance with the KTHNY scenario. This phase is characterized by long-ranged $6-$fold orientational and quasi long-ranged positional order, occasionally disrupted by neutral pairs of thermally activated dislocations. The solid and isotropic liquid phases are, in turn, separated by a hexatic phase, for packing fractions in the range $0.716 < \Phi < 0.720$, while for $0.700 < \Phi < 0.716$ phase coexistence between the hexatic and isotropic liquid phases is observed. 

The snapshots in Figs.~\ref{fig:S3}a-c illustrate the typical configurations found in three of the aforementioned regimes: solid phase (Fig.~\ref{fig:S3}a), hexatic-isotropic liquid coexistence (Fig.~\ref{fig:S3}b) and isotropic liquid (Fig.~\ref{fig:S3}c). As in the main text of the article, red and blue dots denote positive (red) and negative (blue) elementary defects, which in this specific case corresponds to particles with five and seven neighbors respectively. A full computational study of the melting scenario in PBPs was presented by Bernard and Krauth$^{\href{https://doi.org/10.1103/PhysRevLett.107.155704}{25}}$. Figs.~\ref{fig:S3}d-f, on the other hand, show a comparison between the probability distribution of $|\Psi_{p}^{\rm bond}|$ -- i.e. Eq.~\eqref{eq:Psi_bond} - and $|\gamma_{p}|$ -- i.e. Eq.~\eqref{eq:gamma} -- computed respectively from the Delaunay triangulations of the particles and its associated Voronoi diagram. For $p=2$, these two distributions do not exhibit any significant difference as they are both peaked around $0.1$, irrespectively of the state of the system (solid, liquid, or in-between), thus signalling the absence of nematic order at small length scales, as expected for systems of disks. For $p=6$, on the other hand, both distributions peak at around $0.85$ and progressively spread over the entire unit interval as the system is driven toward the liquid phase. Figs.~\ref{fig:S3}g-i show a comparison between the magnitudes of the shape parameter $\Gamma_{p}$ and the bond orientational order parameter $\Psi_{p}^{\rm bond}$ for increasing length scale $R$ when $p=2,\,6$. In all three scenarios, the $R^{-1}$ decay embodies the absence of nematic order at any scale. This is because, regardless of the specific $p$ value, in a collection of $N$ randomly oriented bonds 
\begin{equation}
|\Psi_{p}^{\rm bond}| \sim N^{-1/2}\;,
\end{equation}
while $N\sim R^{d}$. Thus $|\Psi_{p}^{\rm bond}| \sim R^{-1}$ (i.e. $\eta_{p}=2$) for $d=2$. Conversely, both magnitudes transition from finite, in the solid phase, to power law decaying upon melting. As the system becomes progressively more disordered the exponent $\eta_{6}$ increases until attaining the universal value $\eta_{6}=1/4$ (black line) at the solid-hexatic coexistence regime. In the isotropic liquid phase, on the other hand, $\eta_{p}\approx 2$. Finally, the same behavior is reflected, as expected, by the correlation functions, Figs.~\ref{fig:S3}j-l. 

We notice that, irrespectively of the state of the system, the exponent $\eta_{p}$ is the same for both the shape function $\gamma_{p}$ and the bond orientation function $\psi_{p}^{\rm bond}$, while the systematically smaller magnitude of $\Gamma_{p}$ compared to $\Psi_{p}^{\rm bond}$ -- or, equivalently, that of their associated correlation functions -- originates from the fact that $0\le |\gamma_{p}|\le 1$ quantifies the resemblance of a cell and a regular $p-$sided polygon, whereas $|\psi_{p}^{\rm bond}|=1$ by construction. From these observations we conclude that the scaling behavior of the magnitude of shape parameter $\Gamma_{p}$ introduced in the main text is in fact a good indicator of $p-$atic order in planar assemblies of polygonal cells. We also stress that, while in this example the configuration of the polygonal cells is related to that of the bond network by the well known Voronoi-Delaunay duality, the same does not occur in cell monolayers, as in this case cells are not Voronoi polygons. The shape function $\gamma_{p}$ and its scale-dependent shape parameter $\Gamma_{p}$ are, therefore, the only suitable observables to quantify orientational order in these systems.

\section{\label{sec:g_tensor}The $p-$fold shape tensor}

\subsection{\label{sec:2a}Definition and basic properties}

In this supplementary Section we explicit the relation between the $p-$fold shape tensor $\bm{G}_{p}$, Eq.~(1), and the shape function parameter $\gamma_{p}$, given in Eq.~\eqref{eq:gamma}. To build up intuition, we start from observing that the standard rank$-2$ shape tensor for a $V-$sided polygon, is given by
\begin{equation}\label{eq:shape_tensor}
\bm{S} = \frac{1}{V}\sum_{v=1}^{V} \bm{r}_{v} \otimes \bm{r}_{v}\;,
\end{equation}
where $\bm{r}_{v}$ represents once again the coordinate of the $v-$th vertex with respect to the center of mass of the cell$^{\href{https://doi.org/10.1007/s10035-003-0126-x}{15},\,\href{http://dx.doi.org/%2010.1007/s10035-003-0127-9}{16}}$. The spectral theorem allows expressing $\bm{S}$, as well as any other symmetric tensor, in terms of two irreducible components, one diagonal and the other traceless:
\begin{equation}\label{DecompositionG2}
\bm{S} = \bar{\lambda}\,\mathbb{1}+\Delta\lambda\left(\bm{e}_{1}\otimes\bm{e}_{1}-\frac{1}{2}\,\mathbb{1}\right)\;,
\end{equation}
where we have set
\[
\bar{\lambda} = \frac{\lambda_{1}+\lambda_{2}}{2}\;,\qquad
\Delta\lambda = \lambda_{1}-\lambda_{2}\;,
\]
with $\lambda_{1}>\lambda_{2}$ the two eigenvalues of $\bm{S}$, $\bm{e}_{1}=\cos\vartheta\,\bm{e}_{x}+\sin\vartheta\,\bm{e}_{y}$ the unit eigenvector associated with the largest eigenvalue $\lambda_{1}$ and $\mathbb{1}$ the rank$-2$ identity tensor. The two terms in Eq.~\eqref{DecompositionG2} entail information about the polygon's size and anisotropy. The latter property can be further highlighted by introducing the tensor
\begin{equation}\label{eq:g2}
\bm{G}_{2} = \frac{\traceless{\bm{S}}}{\Delta_{2}} = \frac{\Delta\lambda}{\Delta_{2}}\,\traceless{\bm{e}_{1}^{\otimes 2}}\;,
\end{equation}	
where $\Delta_{2}=\sum_{v=1}^{V}|\bm{r}_{v}|^{2}$, the operator $\traceless{\cdots}$ has the effect of rendering its argument traceless and symmetric$^{\href{https://doi.org/10.1103/PhysRevLett.129.067801}{20},\,\href{http://dx.doi.org/10.1103/PhysRevE.106.024701}{21}}$ and the $(\cdots)^{\otimes p}$ implies a $p-$fold tensorial product of the argument with itself: i.e.
\begin{equation}
\bm{e}_{1}^{\otimes p}=\underbrace{\bm{e}_{1} \otimes \bm{e}_{1} \otimes \dots \otimes \bm{e}_{1}}_{p\,{\rm times}}\;.	
\end{equation}
In two dimensions, the tensor $\bm{G}_{2}$ has only two linearly independent components and expressing it in the basis $\{\bm{e}_{x},\bm{e}_{y}\}$ readily gives
\begin{equation}
\bm{G}_{2} = \frac{\Delta\lambda}{2\Delta_{2}}\,
\begin{bmatrix}
\cos 2\vartheta & \sin 2\vartheta \\[5pt]
\sin 2\vartheta &-\cos 2\vartheta
\end{bmatrix}\;.
\end{equation}
Furthermore, explicitly diagonalizing Eq.~\eqref{eq:shape_tensor} gives
\begin{subequations}\label{eq:lambda_theta}
\begin{gather}
\vartheta =\frac{1}{2}\,\arctan\left(\frac{\sum_{v=1}^{V}{|\bm{r}_{v}|^{2}\sin{2\phi_{v}}}}{\sum_{v=1}^{V}{|\bm{r}_{v}|^{2}\cos{2\phi_{v}}}}\right)\;,\\[10pt]
\Delta \lambda= \sqrt{\left(\sum_{v=1}^{V}{|\bm{r}_{v}|^{2}\sin{2\phi_{v}}}\right)^2+\left(\sum_{v=1}^{V}{|\bm{r}_{v}|^{2}\cos{2\phi_{v}}}\right)^2}\;,
\end{gather}
\end{subequations}
where $\phi_{v}=\arctan(y_{v}/x_{v})$ denotes the angular position of the $v-$th vertex with respect of the centre of mass (see Fig.~1c in the main text). This construction implies that all components of the tensor $\bm{G}_{2}$ are proportional to either the real or imaginary part of the complex function
\begin{equation}
\gamma_{2} =\frac{1}{\Delta_{2}}\,\sum_{v=1}^{V}{|\bm{r}_{v}|^{2}e^{2i\phi_{v}}} = \frac{\Delta\lambda}{\Delta_{2}}\,e^{2i\vartheta}\;,
\end{equation}
so that
\[
|\gamma_{2}| = \frac{\Delta\lambda}{\Delta_{2}}\;,\qquad 
\vartheta = \frac{\Arg(\gamma_{2})}{2}\;.
\]
Now, the same construction can be carried out for a generic rank$-p$ shape tensor, by defining
\begin{equation}
\bm{G}_{p} = \frac{1}{\Delta_p} \bigg\llbracket\sum_{v=1}^{V}\bm{r}_{v}^{\otimes p}\bigg\rrbracket\;,
\label{eqn:pRankTensor}
\end{equation}
where $\Delta_{p}=\sum_{v=1}^{V}|\bm{r}_{v}|^{p}$. As for the rank$-2$ tensor defined in Eq.~\eqref{eq:g2}, this tensor has only two linearly independent components, that are
\begin{subequations}\label{eq:g1_g2}
\begin{gather}
g_{1}=G_{p,xx\cdots\,x}=\frac{1}{2^{p-1}\Delta_{p}}\sum_{v=1}^{V}{|\bm{r}_{v}|^{p}\cos{(p\phi_{v})}}\;,\\
g_{2}=G_{p,xx\cdots\,y}=\frac{1}{2^{p-1}\Delta_{p}}\sum_{v=1}^{V}{|\bm{r}_{v}|^{p}\sin{(p\phi_{v})}}\;,
\end{gather}
\end{subequations}
and can be cast as in Eq.~\eqref{eq:g2}, that is
\begin{equation}\label{eq:gp}
\bm{G}_{p} = \frac{\Delta\lambda_{p}}{\Delta_{p}}\,\traceless{\bm{e}^{\otimes p}}\;,
\end{equation}
where the positive scalar $\Delta\lambda_{p}$ and the unit vector $\bm{e}=\cos\vartheta\,\bm{e}_{x}+\sin\vartheta\,\bm{e}_{y}$ are analogous to the difference $\lambda_{1}-\lambda_{2}$, quantifying the anisotropy of the polygon, and the eigenvector $\bm{e}_{1}$ associated with the largest eigenvalue. This problem ultimately relies on a generalization of the spectral theorem for tensors whose rank is larger than two. A possible strategy to achieve such a generalization was proposed by Virga in the context of rank$-3$ tensors$^{\href{https://doi.org/10.1140/epje/i2015-15063-x}{22}}$. This consists of defining $\vartheta$ as the inclination of a $p-$legged star oriented in a way that maximizes the probability of finding a vertex of the polygon in the direction of either one of the legs. The latter task is equivalent to solving the system of equations
\begin{equation}\label{eq:eigenvalue_problem}
\bm{G}_{p} \odot \bm{e}^{\otimes p-1}= \frac{\Delta\lambda_{p}}{\Delta_{p}}\,\bm{e}\;,
\end{equation}
where $\odot$ denotes a contraction of all matching indices of the two tensors on the left hand side. After some lengthy calculations, partially summarized in Sec.~\ref{sec:2}, one finds
\begin{subequations}\label{eq:lambdap_thetap}
\begin{gather}
\vartheta=\frac{1}{p}\arctan\left(\frac{\sum_{v=1}^{V}{|\bm{r}_{v}|^{p}\sin{(p\phi_{v})}}}{\sum_{v=1}^{V}{|\bm{r}_{v}|^{p}\cos{(p\phi_{v})}}}\right)\;,\\[10pt]
\Delta\lambda_{p}=\sqrt{\left(\sum_{v=1}^{V}{|\bm{r}_{v}|^{p}\cos{(p\phi_{v})}}\right)^2+\left(\sum_{v=1}^{V}{|\bm{r}_{v}|^{p}\sin{(p\phi_{v})}}\right)^2}\;.
\end{gather}
\end{subequations}
As in the case of the rank$-2$ shape tensor, one can then express all components of $\bm{G}_{p}$ in terms of the real and imaginary parts of the $p-$fold complex function
\begin{equation}
\gamma_{p} = \frac{1}{\Delta_{p}} \sum_{v=1}^{V}|\bm{r}_{v}|^{p}e^{ip\phi_{v}}=\frac{\Delta\lambda_{p}}{\Delta_p}\,e^{ip\vartheta}\;,
\end{equation}
so that
\[
|\gamma_{p}|=\frac{\Delta\lambda_{p}}{\Delta_p}\;,\qquad 
\vartheta = \frac{\Arg(\gamma_{p})}{p}\;.
\]

\subsection{\label{sec:2}Derivation of Eqs.~\eqref{eq:lambdap_thetap}}

For sake of completeness, here we elaborate on the solution of Eq.~\eqref{eq:eigenvalue_problem}, leading to Eqs.~\eqref{eq:lambdap_thetap}. The strategy, pioneered in Ref.$^{\href{https://doi.org/10.1140/epje/i2015-15063-x}{22}}$, consists of mapping the diagonalization of a rank$-p$ tensor to an optimization problem where $\Delta\lambda_{p}\in\mathbb{R}$ is the Lagrange multiplier subjected to the constraint $|\bm{e}|^{2}=e_{x}^{2}+e_{y}^{2}=1$. This task requires computing the tensorial power $\bm{e}^{\otimes p-1}$, which, in turn, amounts to constructing all possible order$-(p-1)$ products of $e_{x}$ and $e_{y}$. The latter is facilitated by the fact that, as previously stated, the two-dimensional tensor $\bm{G}_{p}$ has only two linearly independent components, proportional to the functions $g_{1}$ and $g_{2}$ introduced in Eqs.~\eqref{eq:g1_g2}. In particular, depending on whether the number of $y-$indices of the generic element $G_{i_{1}i_{2}\cdots\,i_{p}}$, with $i_{p}=\{x,y\}$, is even or odd, the element is proportional to $g_{1}$ and $g_{2}$ respectively. Taken together, the aforementioned considerations result into the following expressions for the components of the $\bm{e}$ vector:
\begin{subequations}\label{eq:longcombexpression}
\begin{gather}
\dfrac{\Delta\lambda_{p}}{\Delta_p} e_{x} 
= g_{1}\left[\sum_{k\in\text{even}}{(-1)^{\frac{k}{2}}\binom{p-1}{k}e_{x}^{p-1-k}e_{y}^{k}}\right]
+ g_{2}\left[\sum_{k\in\text{odd}}{(-1)^{\frac{k-1}{2}}\binom{p-1}{k}e_{x}^{p-1-k}e_{y}^{k}}\right]\;,\\
\dfrac{\Delta\lambda_{p}}{\Delta_p} e_{y} 
= g_{2}\left[\sum_{k\in\text{even}}{(-1)^{\frac{k}{2}}\binom{p-1}{k}e_{x}^{p-1-k}e_{y}^{k}}\right]
- g_{1}\left[\sum_{k\in\text{odd}}{(-1)^{\frac{k-1}{2}}\binom{p-1}{k}e_{x}^{p-1-k}e_{y}^{k}}\right]\;.
\end{gather}
\end{subequations}
Despite their apparently complexity, these equations can be considerably simplified leading to
\begin{subequations}\label{eq:shortcombexpression}
\begin{gather}
\dfrac{\Delta\lambda_{p}}{\Delta_p} \cos{\vartheta} = g_{1}\cos{\left[\left(p-1\right)\vartheta\right]}+g_{2}\sin{\left[\left(p-1\right)\vartheta\right]}\;,\\[10pt]
\dfrac{\Delta\lambda_{p}}{\Delta_p} \sin{\vartheta} = g_{2}\cos{\left[\left(p-1\right)\vartheta\right]}-g_{1}\sin{\left[\left(p-1\right)\vartheta\right]}\;.
\end{gather}
\end{subequations}
If $g_{1}=0$, Eqs.~\eqref{eq:shortcombexpression} reduces to 
\begin{equation}
\cot{\vartheta}=\tan\left[\left(p-1\right)\vartheta\right]\;,
\end{equation}
which has $2\,p$ solutions in the range $0\le \vartheta < 2\pi$ given by 
\begin{equation}\label{eq:theta_sol_1}
\vartheta^{(k)}=\frac{2k+1}{p}\,\pi\;,\qquad k=0,\,1\ldots\,2p-1\;.
\end{equation}
Conversely, when $g_{1}\neq0$, setting $\varrho = g_{2}/g_{1}$ and solving Eqs.~\eqref{eq:shortcombexpression} with respect to $\vartheta$ gives 
\begin{equation}
\cot{\vartheta} = 
\frac
{\cos{\left[\left(p-1\right)\vartheta\right]}+\varrho\sin{\left[\left(p-1\right)\vartheta\right]}}
{\varrho\cos{\left[\left(p-1\right)\vartheta\right]}-\sin{\left[\left(p-1\right)\vartheta\right]}}\;,
\end{equation}
from which one can readily find
\begin{equation}
\varrho =\tan p\vartheta\;,
\end{equation}
whose solution is given by
\begin{equation}\label{eq:theta_sol_2}
\vartheta^{(k)}=\frac{\arctan{\varrho} + k\pi}{p}\;,\qquad k=0,\,1\ldots 2p-1\;,
\end{equation}
thereby completing the derivation of Eq.~(\ref{eq:lambdap_thetap}a). To compute $\Delta\lambda_{p}$ one can use again Eqs.~(\ref{eq:shortcombexpression}) and express $g_{1}$ and $g_{2}$ in terms of coordinates. This gives
\begin{equation}  
\left|\Delta\lambda_{p}^{(k)}\right|
= \left|g_{1}\right|\sqrt{1+\varrho^{2}}\,
= \frac{1}{2^{p-1}}\,\sqrt{\left(\sum_{v=1}^{V}{|\bm{r}_{v}|^p\cos{(p\phi_{v})}}\right)^{2}
+ \left(\sum_{v=1}^{V}{|\bm{r}_{v}|^{p}\sin{(p\phi_{v})}}\right)^2}\;.
\end{equation}
Note that, because of the periodicity of $\vartheta^{(k)}$, then $\Delta\lambda_{p}^{(k)}=-\lambda_{p}^{(k+1)}$, whereas the sign of $\lambda_{p}^{(0)}$ depends on $\varrho$ and $g_{1}$. Finally, to cast the tensor $\bm{G}_{p}$ in the form given in Eq.~\eqref{eq:gp}, one can write $g_{1}=\Delta\lambda_{p}\hat{g}_{1}$ and $g_{2}=\Delta\lambda_{p}\hat{g}_{2}$ where
\[
\hat{g}_{1}=\frac{1}{2^{p-1}\Delta_p}\,\cos{p\vartheta}\;,\qquad
\hat{g}_{2}=\frac{1}{2^{p-1} \Delta_p}\,\sin{p\vartheta}\;,
\]
are the two independent components of $\traceless{\bm{e}^{\otimes p}}/\Delta_p$. Then, using the expression of $\vartheta$ given in Eqs.~\eqref{eq:theta_sol_1} and \eqref{eq:theta_sol_2}, one obtains
\begin{equation}
\Delta\lambda_{p}=2^{p-1}\left|\Delta\lambda^{(k)}_{p}\right|\;,\qquad k=0,\,1\ldots\,2p-1\;,
\end{equation}
which completes the derivation of Eq. (\ref{eq:lambdap_thetap}b).

\section{Defect representation in $p$-atic liquid crystals}

\begin{figure}[t!]
\centering
{\includegraphics[width=\textwidth]{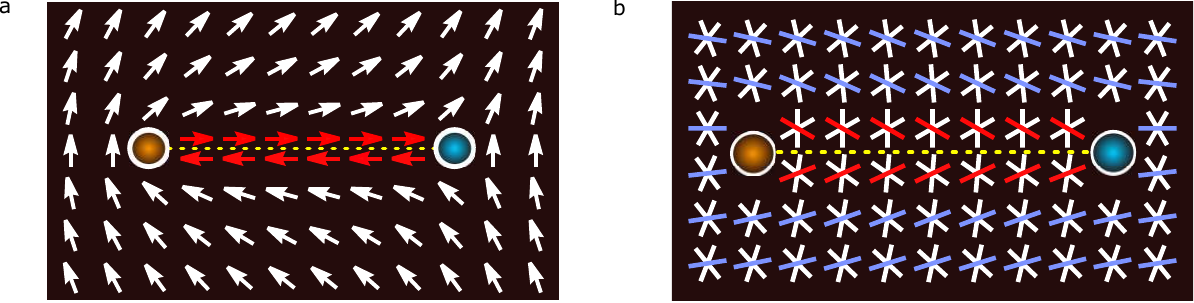}}
\caption{\label{fig:S4}\textbf{Misidentification of singular lines in $p-$atic liquid crystals.}  A mismatch between the defect strength and the underlying $p-$atic symmetry gives rise to unphysical singular lines. (a) In nematics, detecting a pair of $\pm 1/2$ disclinations via a $1-$fold director field (i.e. a standard vector field) yields an intermediate singular line where the director ``jumps'' by $\pi$ discontinuity (red arrows). (b) Analogously, detecting a pair of $\pm 1/6$ hexatic disclinations with a nematic director yields a $\pi/3$ discontinuity (red rods).}
\end{figure}

In two-dimensional liquid crystals, topological defects consists of point-like singularities in the orientational field, that is points where the orientation of the director field is not univocally defined, and can be classified in terms of the winding number $s$ defined in the main text. In liquid crystals with $p-$fold rotational symmetry, the latter is an integer multiple of the elementary winding number $\pm 1/p$. By contrast, it is impossible to correctly describe a defect of winding number $s=\pm 1/p$ in terms of an orientation field with rotational symmetry other than $p-$fold. To substantiate this statement, we consider here the common case of a pair of $\pm 1/2$ disclinations in a nematic liquid crystal ($p=2$), respectively located at positions $\bm{r}_{+}=x_{+}\bm{e}_{x}+y_{+}\bm{e}_{y}$ and $\bm{r}_{-}=x_{-}\bm{e}_{x}+y_{-}\bm{e}_{y}$. The far-field configuration of the phase is given by
\begin{equation}\label{eq:theta_2}
\vartheta =   \frac{1}{2} \left[ \arctan \left(\frac{y-y_+}{x-x_+}\right)-\arctan\left(\frac{y-y_-}{x-x_-}\right)\right]\;.
\end{equation}
In turn, the $2-$fold orientation field can be visualized as the standard headless nematic director $-$ i.e. a $2-$legged star. Now, as illustrated in Fig.~\ref{fig:S4}, attempting to describe the same $2-$fold symmetric configuration with a, say, $1-$fold symmetric orientation filed -- i.e. a standard vector field -- results in a discontinuity of magnitude $\pi$ of the associated phase $\vartheta$ across the $x-$axis. 

The same issue occurs when attempting to describe a pair of $s=\pm n/p$ defects, with $n\in\mathbb{N}$, by means of orientation filed, with lower rotational symmetry. In this case, the far-field configuration of the phase $\vartheta$ is given by
\begin{equation}\label{eq:theta_p}
\vartheta = \frac{n}{p} \left[ \arctan\left(\frac{y-y_+}{x-x_+}\right) - \arctan \left(\frac{y-y_-}{x-x_-}\right) \right], 
\end{equation}
and it can be graphically represented by a $p-$legged star oriented at angles $\vartheta + 2 \pi k/p$, with $k=1,\,2\dots\,p$, so that the functions $\psi_{p}$ and $\gamma_{p}$ (see Sec.~\ref{sec:1}) are continuous everywhere, but at the defect position. Now, let us describe the same configuration in terms of a $q-$fold symmetric function, with $q<p$, which could be graphically represented by a $q-$legged start. Without loss of generality, we set $y_{+}=y_{-}=y_{0}$ and we compute the variation of $\psi_{q}$ while crossing the line the axis $y=y_{0}$ in the region comprised between the two defects ($x_{-} < x < x_{+}$). Since the $q-$legged star associated with the order parameter $\psi_{q}$ is invariant under rotations by $2\pi/q$, the inclination of the leg closer to the $x-$axis undergoes a discontinuity of magnitude
\begin{equation}
|\Delta\vartheta|= 2 \pi \min \left( \dfrac{n}{p},  \dfrac{|p-nq|}{pq} \right)\;.
\end{equation} 
Thus, the field $\vartheta$ is continuous everywhere, but at the defect position ($|\Delta\vartheta| = 2 \pi m/p$ with $m$ any natural number) only when $p=q$ or $nq$ is an integer multiple of $p$. In particular, describing a defect of winding number $s=1/6$ ($n=1$ and $p=6$) by means of a nematic field with $q=2$ would result into a jump of magnitude $|\Delta\vartheta|=\pi/3$ as shown in Fig.~\ref{fig:S4}. The resulting configuration of the nematic director features a singular line connecting defects of opposite strength, as an artifact of this misemployed description. 

\newpage
\FloatBarrier

\section{Extended Data Figures}

\begin{figure}[htbp]
\centering
{\includegraphics[width=0.4\textwidth]{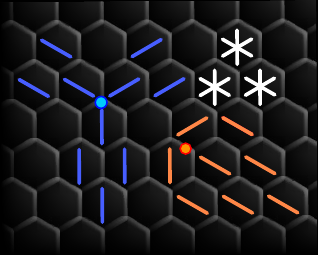}}
\caption{\textbf{Misidentification of nematic defects in a honeycomb lattice.} Because of the $6$-fold symmetry of regular hexagons, there is no well
defined longitudinal direction, thus it is possible to construct a defective configuration, featuring a pair of $\pm1/2$ disclinations, even though the lattice is defect free.}
\label{fig:figE1}
\end{figure}

\begin{figure*}[htbp]
\centering
{\includegraphics[width=\textwidth]{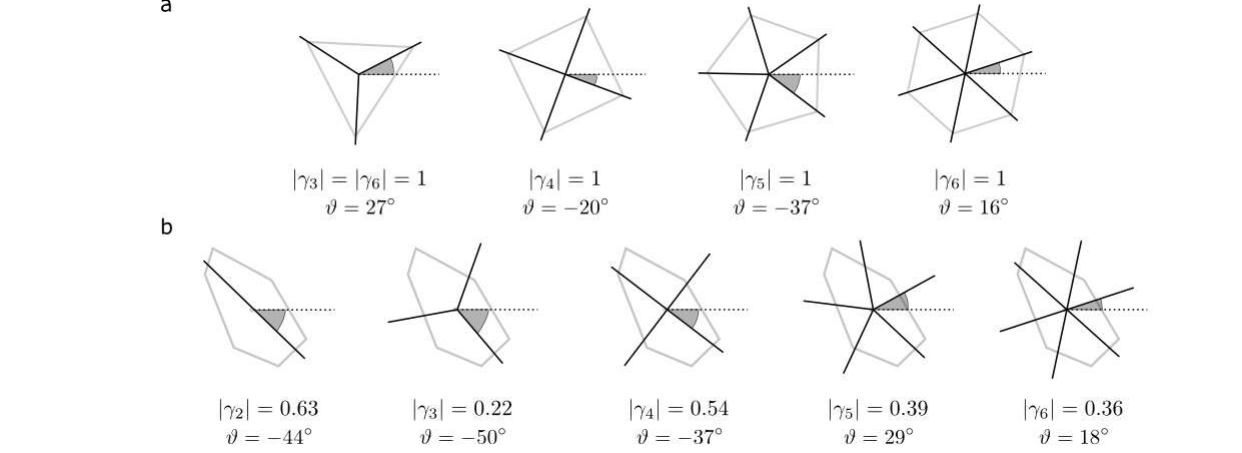}}
\caption{\textbf{Shape function $\gamma_{p}$.} Example of the shape function $\gamma_{p}$ for polygons with different degree of regularity. The black star at the center of each polygon marks its orientation calculated via the phase $\vartheta=\Arg(\gamma_{p})/p$, which is, in turn, highlighted in grey. (a) In the case of regular $V-$sided polygons, $|\gamma_{p}|=\delta_{nV,p}$, with $n\in\mathbb{N}$. (b) For irregular polygons, on the other hand, the magnitude of $\gamma_{p}$ quantifies the resemblance between the polygon and a regular $p-$sided polygon having the same size. Because of its elongated shape, the $6-$sided polygon displayed here has a prominent $2-$fold structure, consistent with the fact that $\gamma_{2}$ is larger in magnitude than any of the other shape functions.}
\label{fig:figE2}
\end{figure*}

\begin{figure}[htbp]
\centering
{\includegraphics[width=\textwidth]{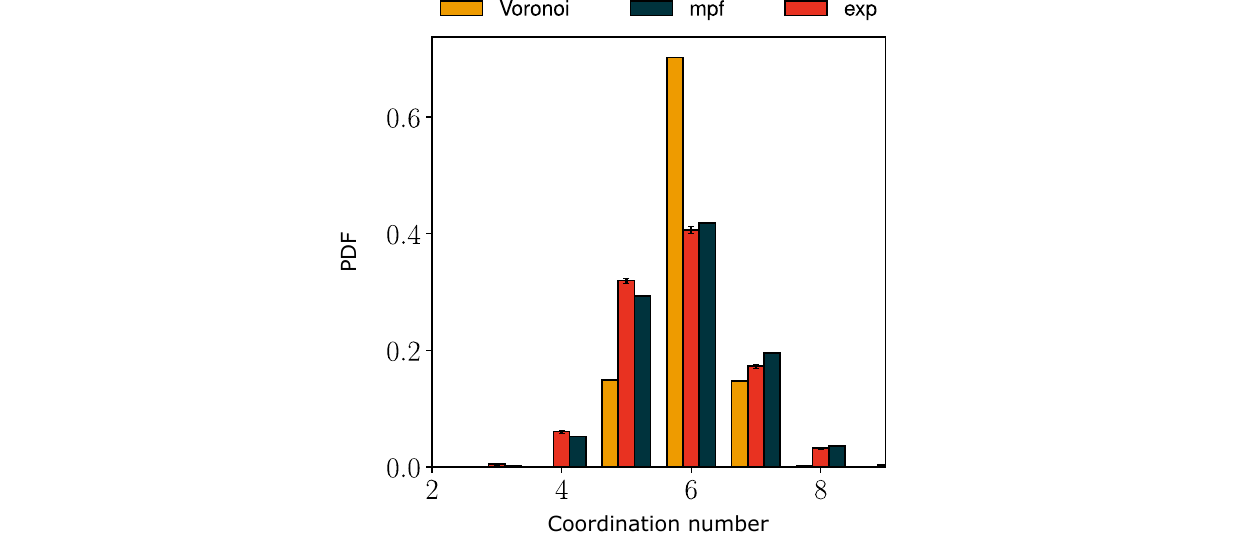}}
\caption{Probability distribution of cell coordination number for experiments and simulations. For the experimental data, the height of the bar represents the mean of $68$ analyzed images. The mean values of the coordination number distributions are $5.8 \pm 0.9$ (mean $\pm$ s.d.) for experiments and $5.9 \pm 0.9$ (mean $\pm$ s.d.) for multiphase field simulations and $6.0 \pm 0.6$ (mean $\pm$ s.d.) for Voronoi simulations.}
\label{fig:figE3}
\end{figure}

\begin{figure}[htbp]
\centering
{\includegraphics[width=\textwidth]{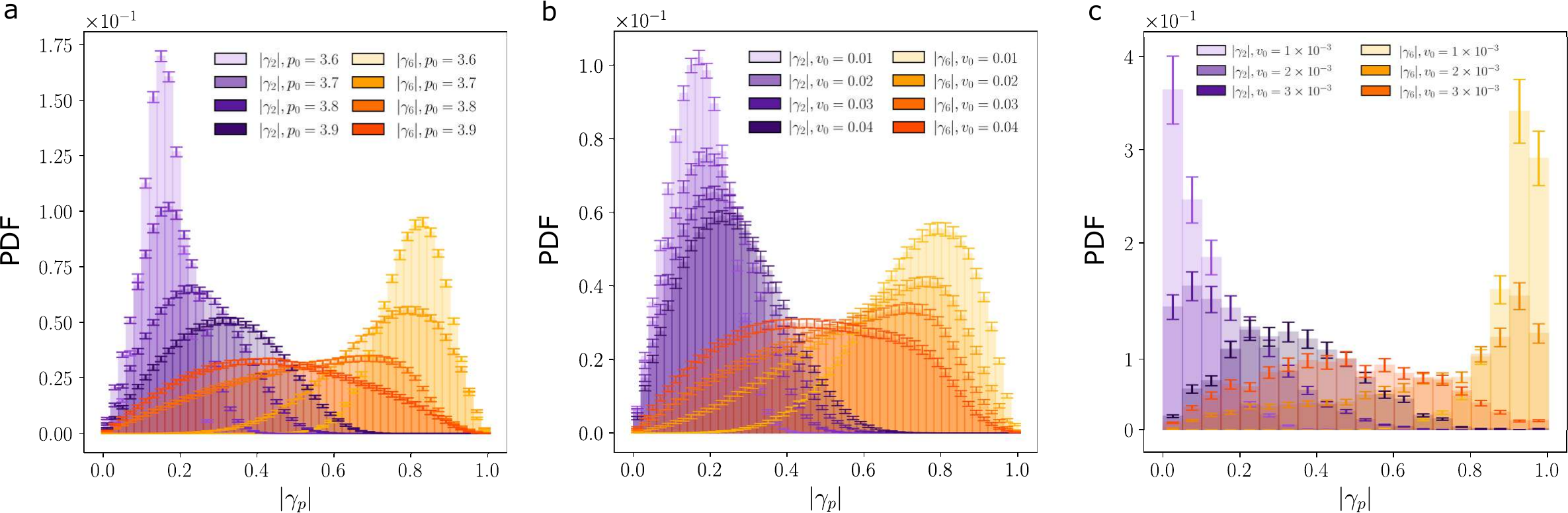}}
\caption{\textbf{Probability distribution of the order parameter's amplitude $\gamma_2$ and $\gamma_6$ for the Voronoi and mpf model as a function of the main simulation parameters.} (a) Probablity distribution of  $|\gamma_p|$, $p=2,6$ for the self-propelled Voronoi model when varying the shape index $p_0$. When increasing $p_0$, the distribution for hexatic shape order parameter flattens and it peaks at lower values. The opposite happens for $p=2$, which turns from being negligible to exhibit a comparable peak value to the one of the hexatic symmetry. These simulations have been performed under the same parameter values provided in Table 1. (b) Same effect when varying the self-propulsion $v_0$. In this case, we lower $p_0=3.7$ in order to capture the melting transition: there is no substantial variation of the profiles when varying $v_0$ at $p_0=3.9$ as the system is deeply in the liquid phase. These observations are coherent with the unjamming transition that has been reported for this particle-based model in Ref.$^{\href{http://dx.doi.org/2010.1103/PhysRevX.6.021011}{30}}$. (c) Probablity distribution of  $|\gamma_p|$, $p=2,6$ for the multiphase field model when varying the velocity of self-propulsion $v_0$. Analogously to the case of the Voronoi model, $|\gamma_6|$ ($|\gamma_2|$) is larger (smaller) at low $v_0$ values.}
\label{fig:figE4}
\end{figure}

\begin{figure}[htbp]
\centering
{\includegraphics[width=\textwidth]{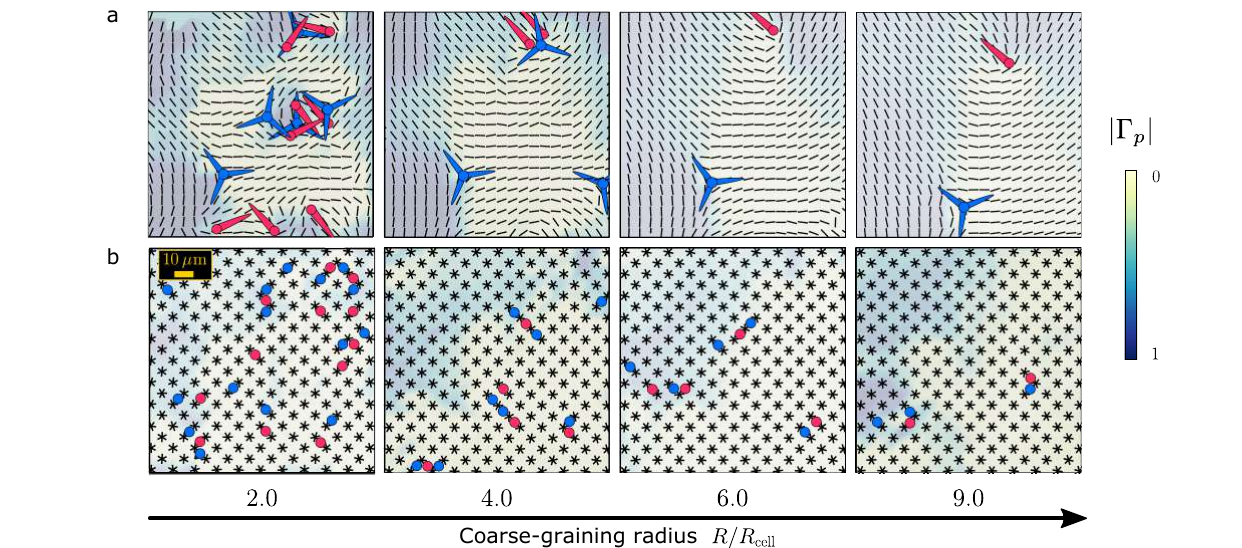}}
\caption{\textbf{Orientational order parameter at different length-scales} Nematic (a) and hexatic (b) shape parameter $\Gamma_2$ and $\Gamma_6$ at different values of the coarse graining radius $R$, expressed in units of the average cell size $R_{\rm cell} = 7.4 \mu$m for the same experimental image as in Fig.~3 of the main text. In both top and bottom panels, positive and negative defects are marked in red and blue respectively ($\pm 1/2$ for nematic and $\pm 1/6	$ for hexatic). The spatial resolution of the shape parameter is the same on both panels, but the number of orientation markers in panel (b) is half of that used in panel (a) for sake of clarity.}
\label{fig:figE5}
\end{figure}

\begin{figure}[htbp]
\centering
{\includegraphics[width=\textwidth]{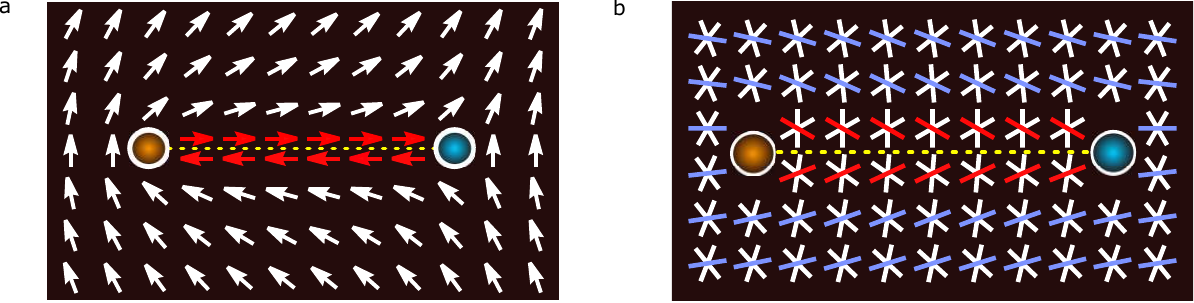}}
\caption{\textbf{Misidentification of singular lines in $p-$atic liquid crystals.}  A mismatch between the defect charge and the symmetry of the $p-$atic liquid crystal gives rise to unphysical singular line. Left (right) panel shows a pair of nematic (hexatic) defects of winding number $s=\pm 1/2$ ($s=\pm 1/6$)}
\label{fig:figE6}
\end{figure}

\begin{figure}
\centering
{\includegraphics[width=1\textwidth]{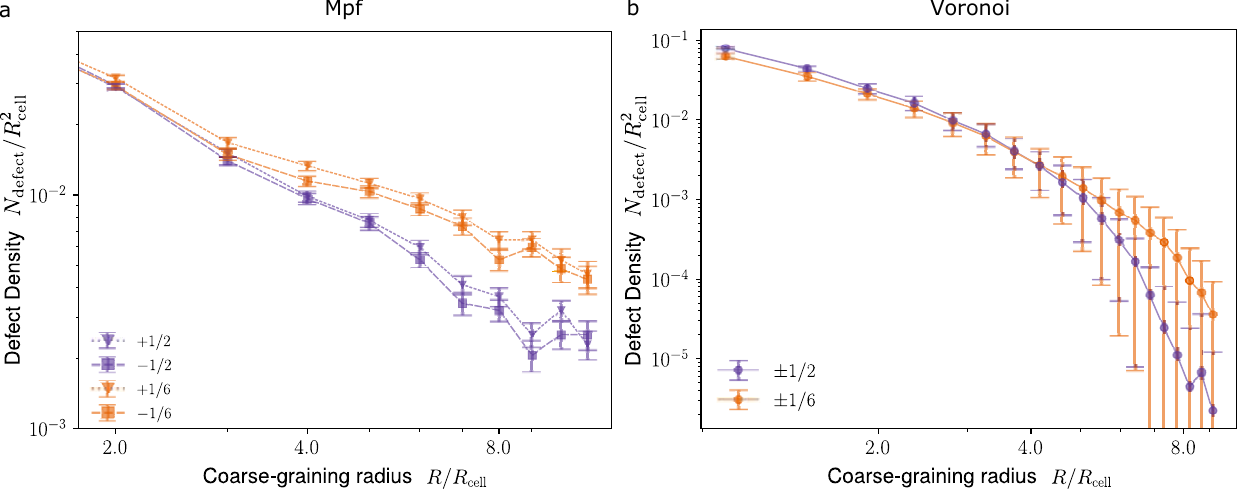}}
\caption{\textbf{Decay of defect density in simulations.} Left and right panels show the decay of the defect density at varying the coarse-graining radius in simulations of the  multiphase field model (left) and Voronoi model (right). The scaling of the multiphase field model data is compatible with that shown in panel d of Fig.~3 in the main text for experimental data. In the right panel, curves of oppositely charged defects overlap exactly.}
\label{fig:figE7}	
\end{figure}

\begin{figure}[htbp]
\centering
{\includegraphics[width=1.\textwidth]{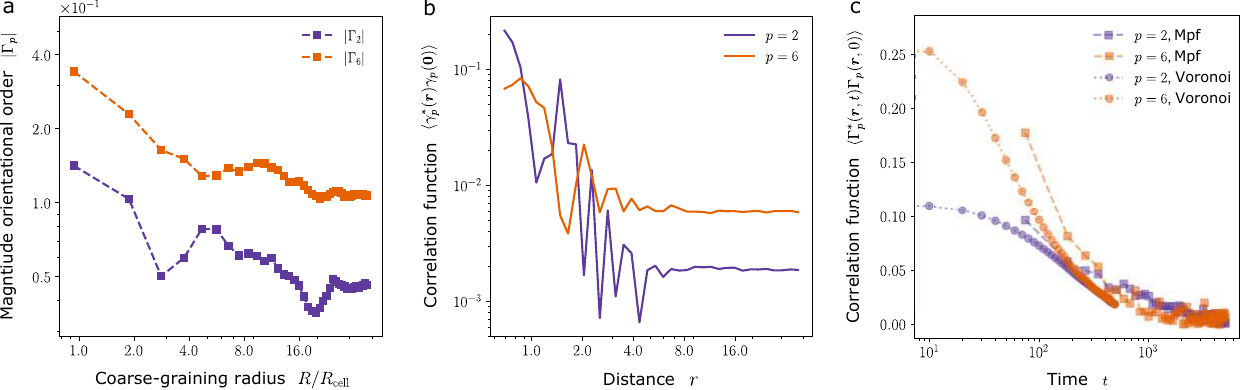}}
\caption{\textbf{Orientational order decay and correlation functions of model epithelia.} (a) Magnitude of
$\Gamma_p$, with $p=2,6$, at varying the coarse-graining radius. As discussed in the main text, because the cells are restricted to be Voronoi polygons, the hexatic symmetry is favored at all length scales and the curves do not crossover. (b) Spatial correlation function of the shape functions $\gamma_2$ and $\gamma_6$. Unlike the correlation functions shown in the main text corresponding to the Multiphase-field model and the experimental results, in this case, the residual hexatic order causes $\left\langle\gamma_6^*(\bm{r})\gamma_6(\bm{0})\right\rangle$ to be greater than $\left\langle\gamma_2^*(\bm{r})\gamma_2(\bm{0})\right\rangle$ at large coarse-graining radius $R$. (c) Decay of time correlation function of the shape parameter $\Gamma_p(\bm{r},t)$ for both Voronoi and Multiphase-field model. These simulations have been performed under the same parameter values provided in Table 1.}
\label{fig:figE8}
\end{figure}

\begin{figure}[htbp]
\centering
{\includegraphics[width=1.\textwidth]{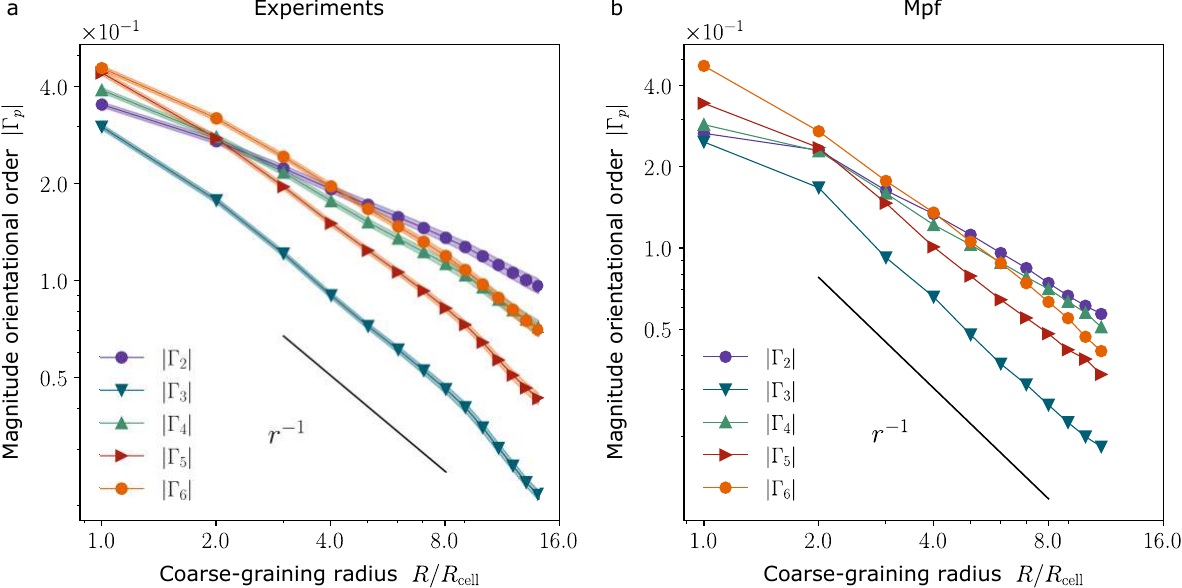}}
\caption{\textbf{Decay of $p-$fold orientational order.} Magnitude of $\Gamma_{p}$, with $p=2,\dots,6$, versus the coarse-graining radius $R$ measured for experimental (a) and numerical Mpf data (b). Orientational order decays with power law behavior $|\Gamma_p| \sim (R/R_{\rm cell})^{-\eta_p/2}$. We notice that $\eta_3,\eta_5 \approx 2$, the critical value corresponding to a disordered state, therefore suggesting $3-$ and $5-$fold symmetry to be significantly suppressed in our experimental and Mpf model tissues. For $p=4$, on the other hand, $\eta_4=1.35 \pm 0.01$ (mean $\pm$ s.e.m.) in experiments and $\eta_4=1.50 \pm 0.03 $ in simulations, thus suggesting the presence of tetratic orientational order. This behavior is, however, the consequence of the degeneracy of the shape funciton illustrated in Section~S1 of the Supplement. Hence, the observation of a residual tetratic order follows from the leading nematic order. Finally, we stress that the largest contribution at small and large scale is given by the hexatic and nematic orded parameter, respectively, for both experimental and simulation data.}
\label{fig:figE9}
\end{figure}